\documentclass[nohyper, 12pt,letterpaper]{JHEP3}
\usepackage[dvips]{epsfig}
\usepackage{amsfonts,amssymb}
\newcommand{\tv}{y}  % Or t. But t can be confused with time?

\newcommand{\IZ}{\mathrm{Z}}

\newcommand{\hs}{\hspace{0.3 cm}}

\newcommand{\mY}{\mathcal{Y}}

\newcommand{\xb}{\bar{x}}
\newcommand{\yb}{\bar{y}}
\newcommand{\zb}{\bar{z}}

\newcommand{\beq}{\begin{equation}}
\newcommand{\eeq}{\end{equation}}
\newcommand{\beqa}{\begin{eqnarray}}
\newcommand{\eeqa}{\end{eqnarray}}

\newcommand{\be}{\begin{equation}}
\newcommand{\ee}{\end{equation}}
\newcommand{\bea}{\begin{eqnarray}}
\newcommand{\eea}{\end{eqnarray}}
\newcommand{\ba}{\begin{eqnarray}}
\newcommand{\ea}{\end{eqnarray}}

\def\a{\alpha}
\def\b{\beta}

\newcommand{\cN}{\mathcal{N}}
\newcommand{\cO}{{\cal O}}

\newcommand{\ads}[1]{\mbox{${AdS}_{#1}$}}
\newcommand{\adss}[2]{\mbox{$AdS_{#1}\times {S}^{#2}$}}

\newcommand{\eg}{{\it e.g.}}

\newcommand{\commentout}[1]{}

\newcommand{\la}{\leftarrow}
\newcommand{\ra}{\rightarrow}
\newcommand{\ua}{\uparrow}
\newcommand{\da}{\downarrow}
\newcommand{\ula}{\nwarrow}
\newcommand{\dla}{\swarrow}

\newcommand{\Rr}{R_{\ra}}
\newcommand{\Rl}{R_{\la}}
\newcommand{\Ru}{R_{\ua}}
\newcommand{\Rd}{R_{\da}}
\newcommand{\Rul}{R_{\ula}}
\newcommand{\Rdl}{R_{\dla}}

\newcommand{\cL}{{\cal L}}

\newcommand{\cY}{\ensuremath{{\cal Y}}}

\newcommand{\half}{\ensuremath{\frac{1}{2}}}

\newcommand{\Pt}{\ensuremath{P_{\tv}}}

\newcommand{\Pps}{\ensuremath{P_\psi}}
\newcommand{\Pph}{\ensuremath{P_\phi}}

\newcommand{\Px}{\ensuremath{P_x}}

\newcommand{\sqt}{\ensuremath{\sin^2 \theta}}
\newcommand{\cqt}{\ensuremath{\cos^2 \theta}}

\newcommand{\amx}{(\alpha-x)}
\newcommand{\bmx}{(\beta-x)}

\newcommand{\dphi}{\dot{\phi}}
\newcommand{\dpsi}{\dot{\psi}}
\newcommand{\dx}{\dot{x}}

\newcommand{\tr}{\mbox{tr}}

\def\lbldef#1#2{\expandafter\gdef\csname #1\endcsname {#2}}

\def\IC{{\mathbb C}}

\def\IZ{{\mathbb Z}}

\newcommand{\Lpqr}{\ensuremath{L^{p,q|r}}}

\newcommand{\opt}{(1+\tv)}
\newcommand{\omt}{(1-\tv)}
\newcommand{\dpsir}{\dot{\psi}_R}
\newcommand{\dt}{\dot{\tv}}
\newcommand{\Ppsir}{P_{\psi_R}}

\title{From Sasaki-Einstein spaces to quivers\\ via BPS geodesics: $\Lpqr$ }

\author{Sergio Benvenuti$^1$, Martin Kruczenski$^2$\\

\vspace{0.6 cm}

\parbox[t]{6in}
{1. Scuola Normale Superiore, Pisa,\\
    and INFN, Sezione di Pisa, Italy.}\\
~\\
{2. Department of Physics, Brandeis University \\
    Waltham, MA 02454, USA.}\\

~\\
\email{sergio.benvenuti@sns.it, martink@brandeis.edu.}
}

\abstract{
The AdS/CFT correspondence between Sasaki-Einstein spaces and quiver
gauge theories is studied from the perspective of massless BPS
geodesics. The recently constructed toric $\Lpqr$ geometries are
considered: we determine the dual superconformal quivers and the
spectrum of BPS mesons. The conformal anomaly is compared with the
volumes of the manifolds. The $U(1)^2_F \times U(1)_R$ global symmetry
quantum numbers of the mesonic operators are successfully matched with
the conserved momenta of the geodesics, providing a test of AdS/CFT
duality. The correspondence between BPS mesons and geodesics allows to
find new precise relations between the two sides of the duality. In
particular the parameters that characterize the geometry are mapped
directly to the parameters used for $a$-maximization in the field theory.

 The analysis simplifies for the special case of the $L^{p,q|q}$
 models, which are shown to correspond to the known
"generalized conifolds". These geometries can break conformal
 invariance through toric deformations of the complex structure.
}

\keywords{AdS/CFT, quiver gauge theory, string theory}
\preprint{\tt{BRX TH-567} \\
 \tt{hep-th/0505206} }

\begin{document}

\section{Introduction} \label{intro}
%%%%%%%%%%%%%%%%%%%%%%%%%%%%%%%%%%%
This paper studies the AdS/CFT correspondence \cite{malda} in the case
of four dimensional $\cN=1$ gauge theories. Type IIB backgrounds of the
form $AdS_5 \times X^5$, where $X^5$ is a Sasaki-Einstein manifold, are dual to a special class of
superconformal gauge theories called quivers. A quiver theory has
product gauge group $\prod SU(N_i)$ and the matter fields transform in
bifundamental representations. For superconformal quivers the two
gravitational central charges $c$ and $a$ are always equal and proportional to the
inverse of the volume of $X^5$. The moduli space of vacua is of the
form $Sym_N(\mathcal{M}_3)$, where $\mathcal{M}_3$ is the Calabi-Yau
cone whose base is the Sasaki-Einstein $X^5$. Another peculiar feature
of these no-flavor theories is the natural existence of long
single-trace mesonic operators, constructed from closed paths in the
quivers, that are dual to semiclassical strings moving in $X^5$.

 A special case of this is ''toric AdS/CFT'': the Calabi-Yau cone
 $\mathcal{M}_3$ is a toric manifold, admitting three $U(1)$
 isometries. In other words $X^5$ is a $T^3$ fibration over a polygon,
 which is drawn on an integer two dimensional lattice. This small set
 of discrete data (that can be encoded in the $U(1)$ charges of the
 Gauged Linear Sigma Model fields describing the toric cone) is enough
 to specify completely the full geometry, and hence also the
 corresponding superconformal gauge theories. Of course, a more
 explicit description, both of the geometries and of the gauge
 theories, is desirable. On the geometric side, in particular, there
 is no known way to determine the toric Sasaki-Einstein metric on
 $X^5$ starting from the toric polygon. On the gauge side, the quiver
 can be thought of as a more explicit description of the
 algebro-geometric structure of the singularity $\mathcal{M}_3$, and
 an algorithm exists \cite{Feng:2000mi}, even though
 it can efficiently handle only small toric polygons; see
 \cite{Morrison:1998cs, Beasley:1999uz, Hanany:2001py, Beasley:2001zp,
 Feng:2000, Feng:2002kk, Feng:2002zw}.

Recently various work has been done in context of toric AdS/CFT, due to
the discovery of infinite sets of explicit Sasaki-Einstein metrics, in
contrast to the previous knowledge of only two examples, namely $S^5$
and $T^{11}$. The latter case was analyzed by Klebanov and Witten
\cite{kw}. The study of these models led to many interesting results.

\cite{Gauntlett:2004zh, Gauntlett:2004yd, Gauntlett:2004hh} found an
infinite set of Sasaki-Einstein metrics on $S^2 \times S^3$, called
$\mY^{p,q}$. These metrics are cohomogeneity one, the isometries being
$SU(2) \times U(1)^2$. One Abelian isometry, generated by the Reeb
vector, is present in any, toric or not, Sasaki-Einstein manifold and
is dual to the $R$-symmetry of the gauge theory. In
\cite{Martelli:2004wu} the toric description was found. The CY cones
are quotients of $\IC^4$; the four GLSM fields have charges $(p+q,
|p-q|; -p, -p)$. The Sasaki-Einstein spaces are smooth precisely when
$p$ and $q$ are coprime.
Recently, a cohomogeneity-two generalization has been found
\cite{Cvetic:2005ft}\cite{MS:2005}, leading to the so called $L^{p,q|r}$
spaces. In fact, the same local metrics on the K\"ahler-Einstein 4d base 
have been found some time ago in the mathematical literature
\cite{ACG:2001}. Moreover, in \cite{ACG:2001} these metrics are shown
to be the most general orthotoric Einstein metrics. The toric data
of the $L^{p,q|r}$ CY cones are a simple generalization of the toric
data of the $\cY^{p,q}$ cones. There are still only four GLSM fields
and a single $U(1)$ action, with integer charges $(p, q; -r, - p - q +
r)$\cite{Cvetic:2005ft}. If $p+q=2 r$ one finds $\mY^{\tilde{p},\tilde{q}}$.

In \cite{Benvenuti:2004dy}, using the toric description of the
singularities, the $\mY^{p,q}$ superconformal quivers have been
constructed; a key role was played by the $SU(2)$ global symmetry:
focusing on toric quivers with this non-Abelian flavor symmetry one is
basically led to the $\mY^{p,q}$ quivers. Various checks of the
  correspondence can be performed \cite{Bertolini:2004xf,
    Benvenuti:2004dy, Herzog:2004tr, Berenstein:2005xa, BK}. Also the
  marginal deformations \cite{Mdef} match \cite{Benvenuti:2005wi, Lunin:2005jy}. A crucial role is played by
  the technique of $a$-maximization \cite{intriligator03}, which
  relies on well established general properties of supersymmetric theories
  \cite{Anselmi:1997am,intriligator03} and is
 thus valid for any 4d SCFT. One, maybe surprising, feature
of the $\mY^{p,q}$ theories, that is important for the present paper is
the following: in the simpler Seiberg dual "phases" of the theories,
namely the toric phases, there is a high degeneracy in the global
symmetry quantum numbers of the bifundamental fields. This can be
understood to be necessary from the AdS perspective: the smallest
dibaryon operators have the same charges (modulo a factor of $N$),
and there are only a very small number of them, since they directly
correspond to the supersymmetric 3-cycles in the geometry.
 
We have recently witnessed a general progress in the algebro-geometric
duality between toric singularities and the gauge theories
living on D3 branes probing the singularities: in \cite{Hanany:2005ve}
Hanany and Kennaway put forward a correspondence between the toric
data and the corresponding quivers. All toric quivers can be
drawn on a torus providing a polygonalization of the torus; every
superpotential term precisely corresponds to a face. The dual graph of
the quiver, the dimer, has a direct physical interpretation in term of
brane setups \cite{Hanany:1996ie}. These setups significantly generalize previously known
constructions \cite{Hanany:1997tb, Hanany:1998, GENCON, Feng:2000}. In \cite{Franco:2005rj} many
features of this picture have been clarified and many examples have
been given. The quiver/dimer $\leftrightarrow$ toric Calabi-Yau's
correspondence is part of a general framework \cite{Okounkov:2003sp}
connecting statistical mechanics of crystals and topological
strings. 

In \cite{BK} the periodic quiver picture was shown to encode
naturally the mesonic operators of the theories. In particular, the
emergence of semiclassical strings directly from paths in the periodic
quivers was discussed: roughly speaking, the direction of a long path
on the quiver is mapped to the position of the string in the toric
base of the Sasaki-Einstein space. Massless BPS geodesic (i.e. point-like massless strings moving only along the
$R$-charge direction) are special cases of these, and turn out to encode a great deal of information about the
structure of the quiver.

The purpose of this paper is to show how the techniques of \cite{BK}
can be extended to generic examples of $\cN=1$ AdS/CFT.

In section \ref{strings} we analyze the geometries, find the angle
associated to the R-charge and determine the properties of massless
point-like strings moving along this direction, that we call BPS geodesics.

Inspired by the results of \cite{Benvenuti:2004dy} and
\cite{Hanany:2005ve}, using the relation between the physical $(p,q)$-webs of five
branes and toric diagrams  \cite{pqwebs, Hanany:2001py}, we then
construct the superconformal quivers associated to the \Lpqr
geometries. An important role in the determination of the global
structure of the quivers is played by the mesonic BPS operators, that
we determine as in \cite{BK}.

The main result of the present paper is a direct comparison between these BPS
mesons and BPS geodesics. As a warm up, in section \ref{pqp} we discuss in detail
this matching for the special cases of $L^{p,q|q}$ spaces. For these
cases the gauge theories are known from \cite{Uranga:1998vf, GENCON} and the analysis
is simpler and somewhat more transparent, both from the gauge side and
from the string side (the quartic equations in these cases become
quadratic, as for the $\cY^{p,q}$s). The end result is a non trivial matching of the
$U(1) \times U(1)$ flavor and $U(1)_R$ conserved charges between BPS
geodesics and BPS mesons.

This comparison is than extended to a general \Lpqr model in section
\ref{generic}. In this section we also provide a direct relation
between the parameters $(\a, \b, x_i)$ characterizing the manifolds
\cite{Cvetic:2005ft} and the parameters used in $a$-maximization
\cite{intriligator03}. 
%These relation can be thought of as an ``off-shell'' check of AdS/CFT correspondence.

\section{Strings moving in the \Lpqr manifold} \label{strings}
 In this section we study point-like massless strings moving in the $\ads{5}\times\Lpqr$ manifold whose
metric is\ \cite{Cvetic:2005ft}:
\beqa
 ds^2&=& - dt^2\, \cosh^2\!\rho + d\rho^2 + \sinh^2\!\rho\, d\Omega_3^2 + ds_{p,q|r}^2 \label{metric1}\\
 ds_{p,q|r}^2 &=& (d\xi+\sigma)^2 + ds_{[4]}^2
\label{metric}
\eeqa
where
\beqa
ds_{[4]}^2 &=& \frac{\rho^2}{4f(x)} dx^2 + \frac{\rho^2}{h(\theta)}d\theta^2 + \frac{f(x)}{\rho^2}
\left(\frac{\sqt}{\alpha}d\phi+\frac{\cqt}{\beta}d\psi\right)^2 \\
 && + \frac{h(\theta)\sqt\cqt}{\rho^2}\left(\frac{\amx}{\alpha}d\phi-\frac{\bmx}{\beta}d\psi\right)^2
\eeqa
and
\beqa
\sigma &=& \frac{\amx\sqt}{\a} d\phi + \frac{\bmx\cqt}{\b}d\psi \\
f(x) &=& x\amx\bmx-\mu \\
\rho^2 &=& h(\theta)-x \\
h(\theta) &=& \alpha\cqt + \beta \sqt
\eeqa
 The geodesics we are interested in sit at the point $\rho=0$ of
 \ads{5} and move in the internal manifold. Therefore in the 
rest of the paper we ignore the \ads{5} part of the background.

 To study such geodesics we first do a change of coordinates and then 
properly identify the angle conjugated to the R-charge. With that
information we find the relation between the conserved charges for
some particular cases of interest which we call extremal geodesics.

\subsection{A change of coordinates}

 In order to facilitate the comparison between BPS geodesics and BPS
mesons in the gauge theory we redefine the variables by
%\footnote{The new
%variable should not be confused with the time in \ads{5} space.}
 $\tv= \cos(2\theta)$, getting
\beq
 ds_{[5]}^2 = \left(d\xi + \sigma \right)^2 + ds_{[4]}^2
\label{metric5}
\eeq
%\beq
% ds_{[5]}^2 = \left(d\xi + \frac{\amx (1-t)}{2 \a} d\phi +
%\frac{\bmx (1+t)}{2 \b} d\psi \right)^2 + ds_{[4]}^2
%\label{metric5}
%\eeq
with \beq \sigma = \frac{\amx\omt}{2 \a} d\phi + \frac{\bmx\opt}{2\b}d\psi \eeq
In such a way the two functions $\sigma_i$ in $d\xi + \sigma_{\phi}
d\phi + \sigma_{\psi} d\psi$ are products of two linear
functions of one coordinate on the toric base ($x$ and $\tv$). This fact arises
naturally when comparing with the gauge theory.

The local K\"ahler-Einstein metric takes the fairly symmetric form
\beqa
ds_{[4]}^2 &=& \frac{\rho^2}{4 f(x)} dx^2 +
\frac{f(x)}{4\rho^2}\left(\frac{\omt}{\a}d\phi+\frac{\opt}{\b}d\psi\right)^2 + \\
 && + \frac{\rho^2}{4 g(\tv) } d\tv^2 +
\frac{g(\tv)}{4 \rho^2}\left(\frac{\amx}{\alpha}d\phi-\frac{\bmx}{\beta}d\psi\right)^2
\eeqa
with
\beqa
f(x) &=& x\amx\bmx-\mu \\
g(\tv) &=& \half (\a +\b - \tv (\a - \b ) )(1-\tv^2) \\
\rho^2 &=& \half (\a + \b - \tv (\a - \b ) - 2 x )
\eeqa
With $\a > \b$ the ranges of the coordinates on the base is $x_1 \leq
x \leq x_2$ and $-1 \leq \tv \leq 1$, where $0 \leq x_1 \leq x_2 \leq
x_3$ are the three roots of $f(x)$. Since $x \leq x_2 \leq \b$ we have
$\rho^2 \geq \half (\a - \b)\omt \geq 0$, for $\tv \leq 1$. Note also that 
$g(\tv)$ is a cubic function of $\tv$ as $f(x)$ is of $x$. 

\subsection{R-charge}

 To be able to compare the results for massless strings with the field theory side we
have to identify the angle conjugate to the R-symmetry. In order to do that, we need to discuss
briefly the computation of the holomorphic 3-form on the Calabi-Yau
cone $\mathcal{M}_3^{p,q|r}$. This is because the
holomorphic three form can be written as $\Omega_{ijk} = \eta^T \Gamma_{ijk} \eta$ with $\eta$ the
covariantly constant spinor. The R-charge rotates the covariantly constant spinor as
$\eta\to e^{\half i \alpha}\eta$ and $\Omega\to e^{i\alpha} \Omega$.

 With the 1-form $\sigma$ defined in the four dimensional base of the
Sasaki-Einstein manifold, we compute its K\"ahler form $k$ and complex structure $J$ as:
\beq
 k = -\half d\sigma, \ \ \ J_a^b = k_{ac} g^{cb} , \ \ P_a^{b} = \half \left(\delta_a^b + i J_a^b\right)
\eeq
where $P_a^b$ projects out the anti holomorphic components. With this projector, we can
find two holomorphic 1-forms $\eta_{1,2}$:
\beqa
\eta_1 &=& \frac{x-\beta}{f(x)} \, dx - \frac{1+\tv}{g(\tv)}\, d\tv + \frac{2i}{\alpha} d\phi \\
\eta_2 &=& \frac{x-\alpha}{f(x)}\, dx + \frac{1-\tv}{g(\tv)}\, d\tv + \frac{2i}{\beta} d\psi \\
\eeqa
 The Sasaki-Einstein metric allows to construct a Calabi-Yau cone
 $\mathcal{M}_3^{p,q|r}$ with metric
\beq
ds^2 = dr^2 + r^2 ds^2_{[5]}
\eeq
In this manifold we introduce a third holomorphic form
\beq
\eta_3 = \frac{dr}{r} + i \left(d\xi + \sigma\right)
\eeq
Now, as argued by Martelli and Sparks in \cite{MS:2005}, the covariantly constant holomorphic 3-form follows as
\beq
\Omega_{[3]} = \sqrt{f(x)g(\tv)}\ e^{i\psi_R}\ r^3\ \eta_1\wedge \eta_2 \wedge \eta_3
\eeq
 The phase $\psi_R$ is conjugated to the $R$-charge and can be determined to be
\beq
\psi_R=3\xi+\phi+\psi
\eeq
from the condition that $\Omega_{[3]}$ should be covariantly constant.

We can therefore rewrite the 5d metric as
\beq
 ds_{[5]}^2 =
%\left(d\xi + \frac{\amx (1-t)}{2 \a} d\phi +
%\frac{\bmx (1+t)}{2 \b} d\psi \right)^2 + ds_{[4]}^2\\
%&&
\left(\frac{d\psi_R}{3} + \left(\frac{\amx\omt}{2 \a} - \frac{1}{3}\right) d\phi +
\left(\frac{\bmx\opt}{2 \b} - \frac{1}{3}\right) d\psi \right)^2 + ds_{[4]}^2
\label{metric5R}
\eeq

\subsection{BPS geodesics}

The action for a massless particle moving along the internal manifold can be written as
\beq
S = \half\left\{(\frac{1}{3}\dpsir+a_1\dphi+a_2\dpsi)^2 + b_1\dx^2+b_2\dt^2+(c_1\dphi+c_2\dpsi)^2+
(d_1\dphi-d_2\dpsi)^2\right\}
\eeq
with the definitions
\beq
a_1 = \frac{\amx\omt}{2\a}-\frac{1}{3} , \ \ a_2 = \frac{\bmx\opt}{2\b}-\frac{1}{3}, \ \ 
b_1 = \frac{\rho^2}{4f(x)}, \ \ b_2 = \frac{\rho^2}{4g(\tv)}
\label{a12def}
\eeq
\beq
c_1 = \frac{\sqrt{f(x)}}{2\rho} \frac{\omt}{\a}, \ \ c_2 = \frac{\sqrt{f(x)}}{2\rho} \frac{\opt}{\b},
\label{c12def}
\eeq
\beq
d_1 = \frac{\sqrt{g(\tv)}}{2\rho} \frac{\amx}{\a}, \ \
d_2 = \frac{\sqrt{g(\tv)}}{2\rho} \frac{\bmx}{\b}
\label{d12def}
\eeq
We can compute the conjugate momenta:
\beqa
\Ppsir &=& \frac{1}{3}\left(\frac{1}{3}\dpsir + a_1 \dphi +a_2 \dpsi\right) \\
\Pt &=& b_2 \dt \\
\Px &=& b_1 \dx \\
\Pph &=& 3 a_1 \Ppsir + c_1 (c_1\dphi+c_2\dpsi) + d_1 (d_1\dphi-d_2\dpsi) \\
\Pps &=& 3 a_2 \Ppsir + c_2 (c_1\dphi+c_2\dpsi) - d_2 (d_1\dphi-d_2\dpsi)
\label{Pdef}
\eeqa
and the Hamiltonian
\beq
H = \frac{9}{2} \Ppsir^2 + \frac{1}{2b_2}\Pt^2 + \frac{1}{2b_1}\Px^2 + \half (\sigma_1^2 +\sigma_2^2)
\eeq
where
\beqa
\sigma_1 &=& (c_1\dphi+c_2\dpsi) \\
\sigma_2 &=& (d_1\dphi-d_2\dpsi)
\eeqa
In terms of the momenta we get
\beqa
\sigma_1 &=& -\frac{-d_2\Pph-d_1\Pps+3(a_1d_2+a_2d_1)\Ppsir}{c_1d_2+c_2d_1} \\
\sigma_2 &=& -\frac{-c_2\Pph+c_1\Pps+3(a_1c_2-a_2c_1)\Ppsir}{c_1d_2+c_2d_1} \\
\eeqa
 Now we consider geodesics that satisfy $\Pt=0$, $\Px=0$ implying that $x=x_0$ and $\tv=\tv_0$
with $x_0$ and $\tv_0$ constant. These constants should be chosen so as
to minimize $H$ as follows form the eq. of motion for $x$ and $\tv$. The minimum
is when $\sigma_1=\sigma_2=0$, which we expect to correspond to a BPS
geodesic. 
%Using that
%\beqa
%a_1d_2+a_2d_1 &=& \frac{\amx\bmx}{\a\b} \\
%a_1c_2-a_2c_1 &=& \frac{\amb}{\a\b}\sqt\cqt
%\eeqa
%the conditions $\sigma_1=\sigma_2=0$ can be written as
%\beqa
%\frac{\amx\bmx}{\a\b} \Pxi &=& \frac{\bmx}{\b} \Pph + \frac{\amx}{\a} \Pps \\
%\frac{\amb}{\a\b} \sqt\cqt \Pxi &=& \frac{\cqt}{\b} \Pph - \frac{\sqt}{\a} \Pps
%\label{geodrel}
%\eeqa
%Alternatively one can see that $\sigma_1=\sigma_2=0$ 
\begin{table}[!h]
\begin{center}
$$\begin{array}{|c|c|c|c|c|} \hline
  &\hs \tv \hs &\hs x \hs&\hs P_{\phi}/\Ppsir \hs &\hs P_{\psi}/\Ppsir \hs\\ \hline\hline

 LD & -1 & x_1 & 3\left(\frac{2}{3} - \frac{x_1}{\a}\right) & - 1 \\\hline

 RU & + 1 & x_2 & - 1 & 3\left(\frac{2}{3} - \frac{x_2}{\b}\right)\\\hline

 LU & + 1 & x_1 &- 1 & 3\left(\frac{2}{3} - \frac{x_1}{\b}\right) \\\hline

 RD & -1 & x_2 & 3\left(\frac{2}{3} - \frac{x_1}{\a}\right)& - 1\\\hline

\end{array}$$
\caption{Values of the flavor charges for the four extremal BPS geodesics.
The names are related to their field theory interpretation}
\label{mesoncharges}
\end{center}
\end{table}

This implies $\dphi=\dpsi=0$ and from
(\ref{Pdef}):
\beq
 \Pph = 3a_1 \Ppsir , \ \ \ \Pps = 3a_2 \Ppsir
\eeq
%which is equivalent to (\ref{geodrel}) after using the definitions (\ref{a12def}).
Using the definitions (\ref{a12def}) we can rewrite this as
\beqa
\frac{\Pph}{\Ppsir} &=& \frac{ \a - 3 \a \tv - 3 x + 3 x \tv}{2 \a} \\
\frac{\Pps}{\Ppsir} &=& \frac{ \b + 3 \b \tv - 3 x - 3 x \tv}{2 \b}
\eeqa
These eqs. give the ratios of the conserved charges as a function of
the $(x,y)$ position of the BPS geodesic. We will be in particular
interested in the values of these ratios at the four vertices of the
rectangle parameterized by $(x,\tv)$. Here the
conserved momenta assume special values that we show in table
\ref{mesoncharges}. Since multiplying two operators the $U(1)$ charges
simply add, it is enough to check the duality between geodesics and
BPS mesons at these four extremal
points, that corresponds to the four vertices of the toric diagrams,
as we see in the next section.
%%%%%%%%%%%%%%%%%%%%%%%%%%%%%%%%%%%%%%%%%%%%%%%%%%%%%%%%%%%%%%%%%%%%%%%%%%%%%
\section{From GLSM charges to quivers and chiral rings: $\Lpqr$}
In this section we pass from the toric data of the \Lpqr singularities
to a representation of the gauge theories\footnote{This problem is addressed 
from the dimer perspective of \cite{Hanany:2005ve, Franco:2005rj} also in \cite{Franco:2005sm}\cite{Butti:2005sw} and \cite{HV:2005}}. Modulo redefinition of the variables, one can assume $p \leq s \leq r \leq q$ (recall
$p+q=r+s$). We also assume that there is no overall common divisor for
the $p,q,r,s$, otherwise one is dealing with Abelian orbifolds. 

\begin{figure}\begin{center}
\epsfig{file=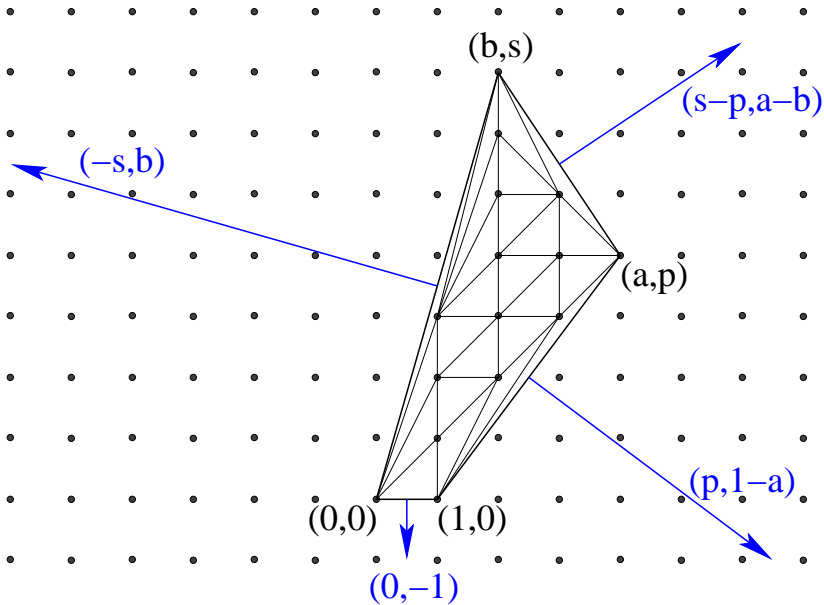, height=8cm}
\caption{Generic toric diagram with four edges. The related $(p,q)$-web is depicted.}
\label{toricdiag}\end{center}
\end{figure}

\begin{figure}\begin{center}
\epsfig{file=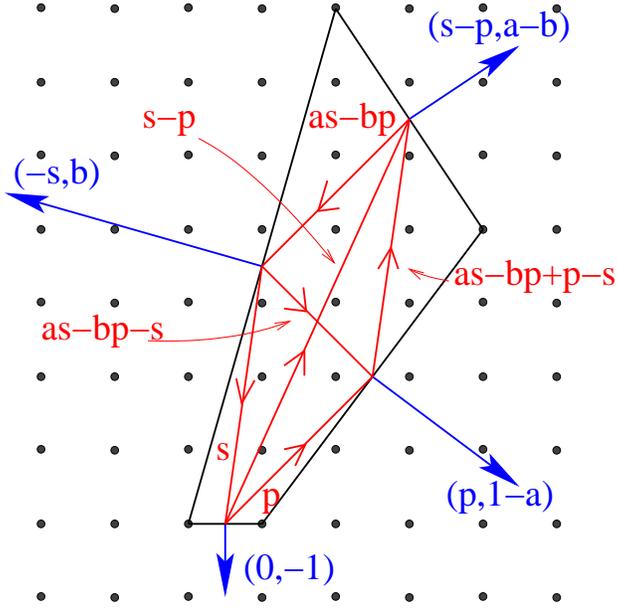, height=8cm}
\caption{Generic toric diagram with four edges. The intersection numbers are computed.}
\label{pqweb}\end{center}
\end{figure}

Since the coordinates on the toric base lie on a polygon
with four edges, it is clear that the toric diagram have four edges as
well\footnote{As we will see from the BPS mesons, if some of the four
pairs $(p,s)$, $(p,r)$, $(q,s)$, $(q,r)$ have a non trivial common
divisor, there are additional points on the boundary of the toric
diagram. For the moment we will assume no non trivial common
divisors, that leads \cite{MS:2005} to smooth geometries.}. Using
$SL(2,\IZ)$ redefinitions of the toric quadrangle, one can assume that
two vertices are in position $(0,0)$ and $(1,0)$. A generic four-edges
toric diagram is reported in figure \ref{toricdiag}.
The natural correspondence between toric diagrams and $(p,q)$-webs of
five branes \cite{pqwebs} is also emphasized. As explained in
\cite{Hanany:2001py}, going into the mirror picture, one can
interpret the number of bifundamental fields of the quiver as
intersection numbers of $3$-cycles in the mirror picture. These
intersection numbers correspond to ``intersections'' of
$(p,q)$-branes, given by the vector product \cite{Hanany:2001py}:
\beq\label{intersect}
\left( \mathcal{C}_i \circ \mathcal{C}_j \right) = p_i q_j - p_j q_i
\eeq
Using this simple formula we compute the $(p,q)$-branes intersection
for our case of interest in figure \ref{pqweb}.

\begin{figure}\begin{center}
\epsfig{file=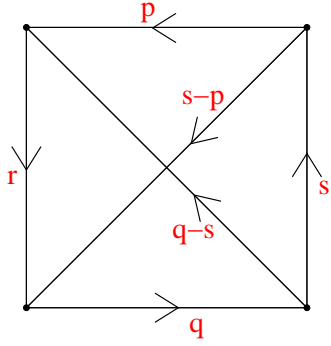, height=4.5cm}
\caption{From fig.\ref{pqweb} we extract the multiplicities of the bi-fundamental fields. We define
$q=as-bp$ and $r=p+q-s$.}\label{foldquiv}\end{center}
\end{figure}

Since intersection numbers can be thought as vector products, they are 
positive when the sine of the angle between vectors is positive. Since the multiplicities 
are positive that allows to determine the direction of the arrows denoting the bifundamental fields.
 It is easy to see that the multiplicities depend only on the
 combination $q=as-pb$ (using $SL(2,\IZ)$ redefinitions, one can
 assume $a \leq p$). It is also convenient
to define $r=p+q-s$ (implying $p+q=r+s$). The resulting bifundamentals together with their multiplicities 
are depicted in figure \ref{foldquiv}. With our definitions, we are
thus describing the toric diagram corresponding to $GLSM$ charges $(p, q; -r, - p - q +
r)$ \cite{Cvetic:2005ft}. $p,q,r,s$ are precisely the multiplicities
of four of the type of the fields that have to appear in the toric
phases of the quiver. According to their direction, we call these four
fields $U \ua$, $D \da$, $L \la$ and $R \ra$.

Figure \ref{foldquiv} should be thought
of as the ``folded'' superconformal quiver. In other words one can
read off form that the number of the various types of fields and the
number of nodes. There are two additional diagonal fields $\ula$ and
$\dla$.\footnote{The picture we are proposing here is expected to
 be valid only for a subset of the toric phases. For examples for the
 $\cY^{p,q}$ in the so called ``double impurity phases'' 
 there are also diagonal fields pointing rightward \cite{Benvenuti:2004wx}.} This representation is very useful in determining the global
symmetry quantum numbers of the bifundamentals. We call the two toric
$U(1)$ symmetries $J_H$ and $J_V$, according to their orientation. The
values are reported in table \ref{charges1}.

Another way of finding the global symmetries from figure
\ref{foldquiv} is to associate a symmetry to every of the four vertices, say $Q_1,
\ldots, Q_4$. The charges are $+1$ for bifundamentals outgoing from
that vertex, $-1$ for ingoing bifundamentals and $0$ for the
others. The sum of these four symmetries vanishes, and a particular
linear combination can be seen to be the baryonic
symmetry.\footnote{In general the number of baryonic symmetries is
equal to $n-3$, where $n$ is the number of external $(p,q)$-branes. In our
picture, for more generic toric diagrams, one constructs immediately
$n$ global Abelian symmetries. The sum is always decoupled, two of them
are the standard toric Abelian isometries, and the remaining $(n-3)$ are
the baryonic symmetries.}

The way to prove that all the field with the same direction have the
same global symmetry quantum number (recall however that they have
different gauge quantum numbers) is simply to impose the vanishing of
the anomalies of these currents for each node.

\begin{table}[!h]
\begin{center}
$$\begin{array}{|c|c|c|c|c|c|c|} \hline
\ \mbox{Field}\ & \ \mbox{Number} \ & R_0 &\ Q_H \ &\ Q_V &\ Q_3 &\ Q_B \\ \hline\hline
R\ \ra & q & 1 & +1 & 0 & 0 & + p \\\hline
L\ \la & p & 1 & -1 & 0 & +1 & + q \\\hline
U\ \ua & s & 0 & 0 & +1 & -1 & - r \\\hline
D\ \da & r & 0 & 0 & -1 & 0 & - s \\\hline
 \ula & q - s & 1 & -1 & +1 & 0 & q - r\\\hline
 \dla & q - r & 1 & -1 & -1 & +1 & q - s\\\hline
\end{array} $$
\caption{Charge assignments for the basic fields. The charge $Q_3$ is redundant but plays a role in the calculations.
$R_0$ is a trial $R$-charge which satisfies the same anomaly cancellation constraints as the actual $R$-charge.}
\label{charges1}
\end{center}
\end{table}

\begin{figure}\begin{center}
\epsfig{file=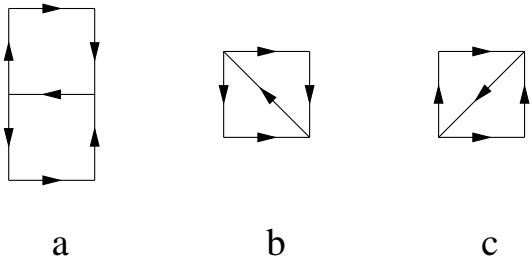, height=3.4cm}
\caption{Building blocks of the toric representation of the quiver.}\label{blocks}\end{center}
\end{figure}

 By studying the possible terms in the superpotential one concludes that the toric representation of the 
quiver can be constructed using the blocks depicted in figure \ref{blocks}. These block should be glued, respecting 
the orientation of the sides, into a fundamental domain with a given number of each type. Finally it should be 
possible to draw the fundamental domain on a torus (equivalently it should be possible to tile the plane with it)
giving rise to identifications between the points in  the boundary of such domain. This is better described with 
an example. In fig.\ref{periodquiv} and table \ref{quivers} we give a construction with $n_a=1$, $n_b=3$, $n_c=1$. 

 In general, the number of blocks of type $(a)$, $(b)$ and $(c)$ are: $n_a = p$, $n_b=q-s$, $n_c=q-r$.
The total number of nodes in the quiver is $p+q$, the total number of bifundamental fields is $p + 3 q$. 
This gives a total number of each type of field in agreement with table \ref{charges1}.

\begin{figure}\begin{center}
\epsfig{file=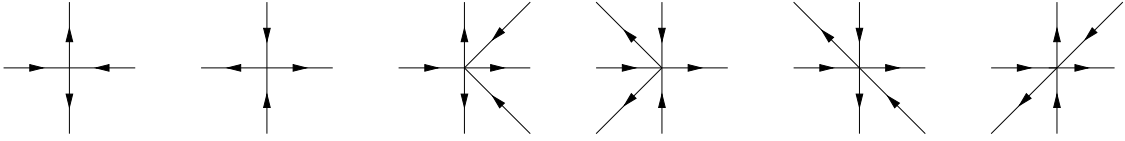, height=1.8cm}
\caption{Vertices (or nodes) appearing in the toric representation of the quiver.}
\label{vertices}\end{center}
\end{figure}

 Gluing these fundamental blocks gives rise to vertices of the type in figure \ref{vertices}. The field
theory interpretation is that each vertex corresponds to an $SU(N)$ gauge group and therefore implies 
anomaly cancellation conditions for the fields ending or emerging from it, or equivalently the beta function 
of the corresponding coupling should be zero.

Also, in the blocks of
figure \ref{blocks}, each square or triangular face corresponds to a term in the superpotential and therefore also implies
a relation between the R-charges of the field surrounding it. Namely, they have to add up to two if the corresponding 
superpotential coupling has zero beta function.

The constraints reduce to the following independent equations
\beqa
\Rr +\Rl +\Ru + \Rd &=& 2 \label{volumesMSY}\\
\Rr + \Rd + \Rul &=& 2 \\
\Ru + \Rdl + \Rr &=& 2
\eeqa
 A simple way to satisfy all this constraints is to assign the following R-charges
$\Ru=\Rd=0$, $\Rl=\Rr=\Rul=\Rdl=1$.
 
However we can add an arbitrary solution of the
homogeneous equations which can be parameterized in terms of three real numbers ($x$, $y$, $z$):
\beqa
\Rr &=& 1 + x \nonumber \\
\Rl &=& 1 - x + z \nonumber \\
\Ru &=& y -z \label{Rcharges}\\
\Rd &=& -y \nonumber\\
\Rul &=& 1 -x + y \nonumber\\
\Rdl &=& 1 -x -y +z \nonumber
\eeqa
It is easily seen that $x$ and $y$ correspond to $Q_H$ and $Q_V$, the $U(1)\times U(1)$ global symmetries of the theory.

On the other hand $z$ can be associated with a $U(1)$ symmetry $Q_3$ in terms of which the baryonic charge is written as
$Q_B = p Q_H + s Q_V + (p + q) Q_3$. The charge assignments are summarized in table \ref{charges1}. There are only three
independent charges. One can use $Q_3$ which is simpler or the baryonic charge
which conveys more physical information since under it all meson operators are neutral.

The global symmetry currents satisfy
$tr(R_0) = tr (J_B) = tr (J_H) = tr (J_V) = 0$.
This is due to the quiver structure of the theory:
\beq
tr(J) = \sum_{f \in fields} J[f] = \half \sum_{i,j \in nodes} J[f_{i,j}] = 0
\eeq
The last term vanish because, for each $i$, $ \sum_{j} J[f_{i,j}] = 0$,
due to the fact that the symmetries are free of anomalies.\footnote{In
  the case of toric quivers, for
  the non-R currents, one can see this by noticing the $tr(J)$ can be
  recast as (one half) the sum over all the faces of the total charges of
  the face, which has to vanish since the superpotential respects the
  symmetry. For the R-symmetry, a similar condition tells that the quiver
  lives on a two-torus \cite{Franco:2005rj}.}

One can also check that $tr(J_B^3)=0$, as has to be the case for a
baryonic symmetry. In our parameterization, the two $U(1)$ symmetries, $Q_H$ and $Q_V$,
also happen to satisfy $tr(J_F^3)=0$. This fact changes by mixing the two flavor currents
between each other or with the baryonic current. The identification of
$J_B$ with the baryonic symmetry comes from the fact, as we will see,
that all mesonic operators (traces of products of bifundamentals) are
uncharged under this symmetry.

\subsection{BPS mesons}
In this subsection we want to determine the BPS mesonic operators of
the theories, that will be compared with the massless BPS geodesics.
The BPS operators that we need to compare with the geodesic are
such that they have maximal $U(1)$ charges (in modulus) for given
R-charge. These are the geodesics that lie precisely on the boundary
of the coordinate rectangle. There are four boundaries, corresponding
to the $4$ supersymmetric $3$-cycles, and four vertices. The
operators that correspond to these four ``corner'' geodesics thus encodes all the
information about BPS geodesics. As a consequence we focus on these
four extremal mesons. It is clear also that these four mesons precisely
correspond to the four $(p,q)$-branes.
\begin{figure}\begin{center}
\epsfig{file=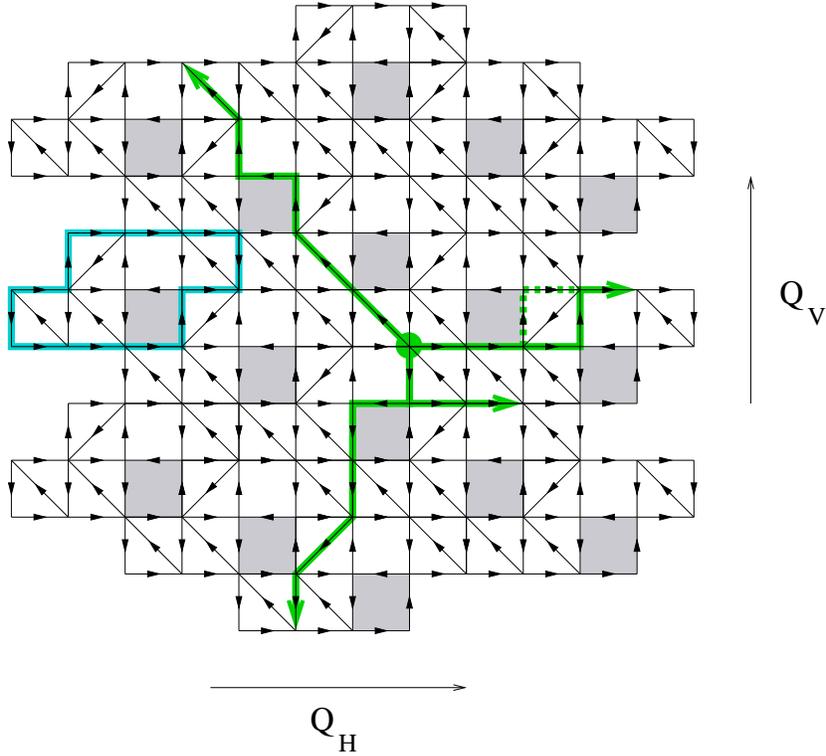, height=10cm}
\caption{Periodic quiver for the $L^{1,5|4}$ model. The four
    ``extremal'' BPS meson and the fundamental domain are
    highlighted. One square is shaded to make the knight
    periodicity easier to appreciate. The dotted line shows an alternative $LU$ (left-up) operator 
with the same charges. The operator in the chiral ring is a linear superposition of the various 
possibilities.}
\label{periodquiv}\end{center}
\end{figure}
The study of the mesons goes further with respect to the folded quiver
picture of the last subsection, and needs a precise understanding of
the way in which the blocks are glued together. If there is an overall
common factor for the four GLSM charges $p,q,-r,-s$, moreover, there
can be various quivers corresponding to the same folded quiver. The
 simplest examples are the two different $\IZ_2$ quotients of the
 conifold. Assuming no overall common divisor there is only one
 quiver; determining this gives also the global charges of our four
 extremal BPS operators. 
\begin{table}[!h]\begin{center}$$\begin{array}{lr} 
 \epsfig{file=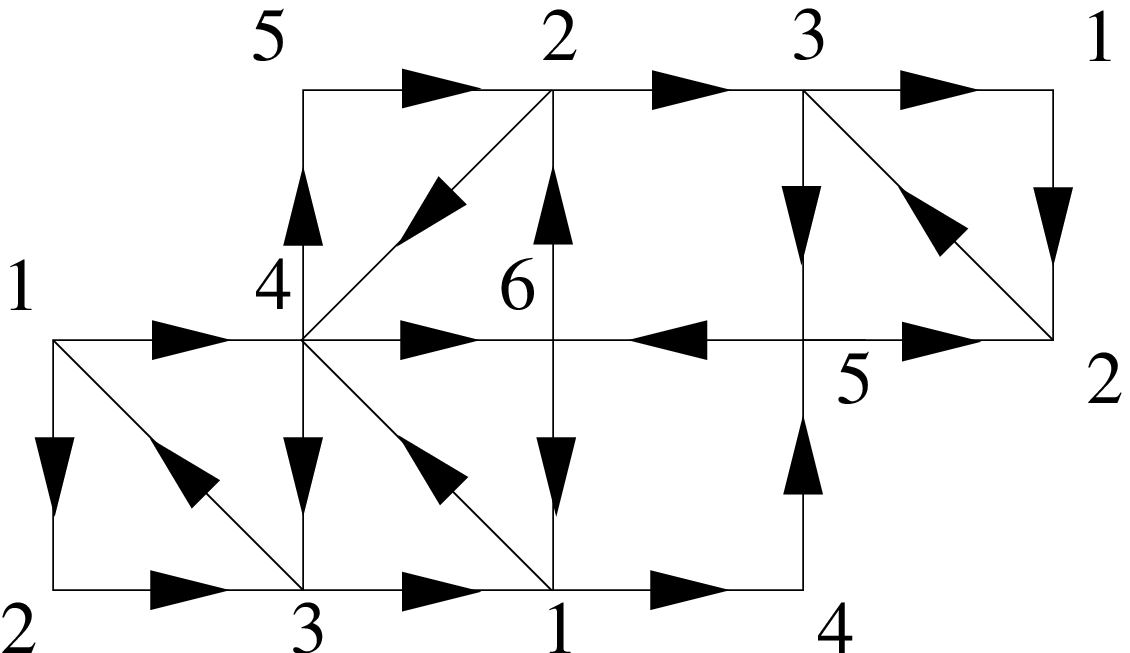, height=3cm} \hspace{1 cm}&
\hspace{1 cm} \epsfig{file=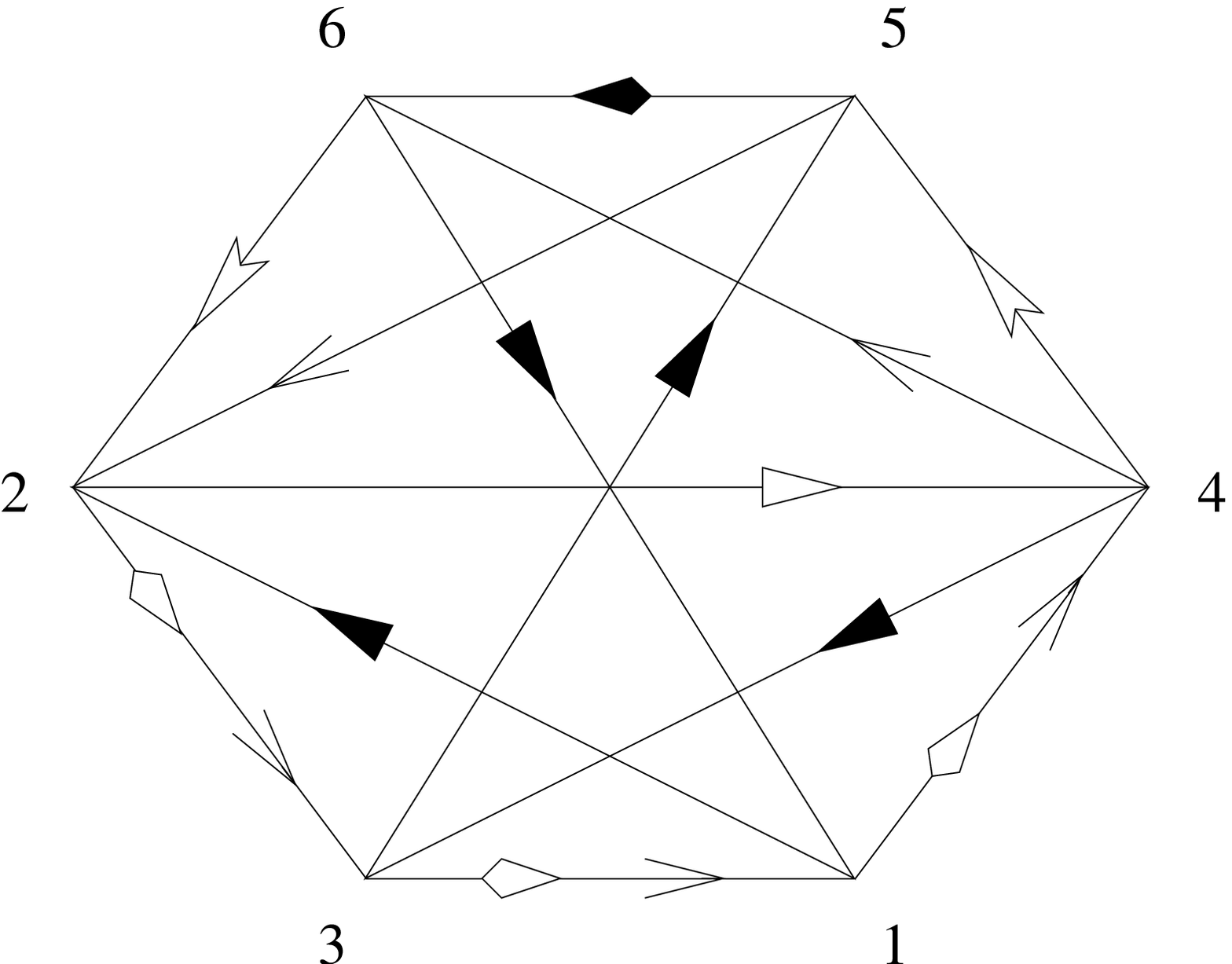, height=3.3cm} 
\end{array} $$
\caption{The fundamental domain of $L^{1,5|4}$ and the traditional
quiver. The numbers identify the six gauge groups and also determine
how the fundamental domain is tiled to cover the plane as in figure \ref{periodquiv}.
Notice that whereas the quiver only contains the information about the matter content
of the theory, the periodic quiver, also tell the superpotential \cite{Hanany:2005ve, Franco:2005rj}.}
\label{quivers}\end{center}\end{table}

\begin{table}[!h]\begin{center}
$$\begin{array}{|c|c|c|c|c|c|} \hline
\ \mbox{Meson}\ & R_0& R &\ Q_H \ &\ Q_V  &\ Q_B \\ \hline\hline
\cO_{LD} &s &-q y + s(1-x+z) & -s & -q  & 0 \\\hline
\cO_{RU} &r & r(1+x)+p(y-z)& +r & +p & 0 \\\hline
\cO_{LU} &r & r (1-x+z) +q (y-z)& -r & +q  & 0 \\\hline
\cO_{RD} &s & s(1+x) -py& +s & -p  & 0 \\\hline
\end{array} $$
\caption{Charge assignments for the four extremal BPS mesons. The
  variables $(x,y,z)$ are taken at the local maximum of the central
  charge $a$.}\label{mcharges1}\end{center}\end{table}

The extremal operator going in the right-up direction, $\cO_{RU}$, for instance, 
is composed  of $\ra$ and $\ua$ fields.  The extremal operator going in the left-up
direction, $\cO_{LU}$ is composed only of $\la$, $\ua$ and $\ula$ fields. Similarly 
for the other two operators. These requirements uniquely determine the four operators.

% There is one $\cO_{LD}$ operator, made of $s$ $L$-fields and $q$ $D$-fields;
%one $\cO_{RU}$ operator, made of $r$ $R$-fields and $p$ $U$-fields,
%one $\cO_{LU}$, made of $r$ $L$-fields and $q$ $U$-fields and finally $\cO_{RD}$, made of $s$ $R$-fields and $p$ $D$-fields.

Without entering into the details, the end result is pretty simple. We assume for the moment that
$\gcd(p,s)=\gcd(p,r)=\gcd(q,s)=\gcd(q,r)=1$ and summarize the general results in table \ref{mcharges1}. 
As an explicit example, in figure \ref{periodquiv} we depict the 
four extremal operators for the particular case of $L^{1,5|4}$.

We can now proceed to understand what happens when there is a non
trivial common divisor between $p$ or $q$ and $r$ or $s$. In these
 cases it is easy to see that the toric diagrams, determined by the
 construction of the previous subsection in term of the GLSM fields,
 have additional points on the edges. This corresponds to additional
 $(p,q)$-legs and reflects as non trivial multiplicities for our
 four BPS mesons. Namely there are $\gcd(s,q)$ $\cO_{LD}$ operators,
 $\gcd(r,p)$ $\cO_{RD}$ operators, and so on. Correspondingly, these
 chiral ring generators become shorter, and their $U(1)$ charges are
 divided by $\gcd(s,q)$ for $\cO_{LD}$, by $\gcd(r,p)$ for $\cO_{RD}$ etc.
This does not affect the ratio between charges. So, for the purpose of comparing
with the geodesics, this fact does play a relevant role.

% In the toric representation of the quiver these are paths which are
% close after we periodically identify the diagram (see fig.\ref{periodquiv} for an example). Therefore, 
%they are gauge invariant.
% Further, here we assume that each pair of the four integers $p,q,r,s$ are
%coprime. If not the operators that we described are wound around the quiver
%several times.

%%%%%%%%%%%%%%%%%%%%%%%%%%%%%%%%%%%%%%%%%%%%%%%%%%%%%%%%%%%%%%%%%%%%%%%
%%%%%%%%%%%%%%%%%%%%%%%%%%%%%%%%%%%%%%%%%%%%%%%%%%%%%%%%%%%%%%%%%%%%%%%
\section{Resolving the strip: $L^{p, q|q}$} \label{pqp}
In this section we focus on a subset of the $\Lpqr$ models that
lies at the opposite boundary with respect to the $Y^{p,q}$ models,
namely $L^{p, q|q}$. In these case one gets the so called generalized
conifolds, studied in detail in \cite{Uranga:1998vf, GENCON}. The corresponding toric
diagrams have no internal points and lie inside the so called strip.

We restrict as usual to $p \leq q$. For $p=q$ we have $\IZ_q$ orbifolds of the
conifold, and for $p=0$ the $\cN=2$ Abelian $\IZ_q$ orbifolds of $S^5$. The $Y^{p,q}$ models can
be thought of as an interpolation between $S^5$ and $T^{11}$ that preserves an $U(2)$
flavor symmetry \cite{Herzog:2004tr}. In the same way the $L^{p,q|q}$ models can be thought of as an
interpolation between $S^5$ and $T^{11}$ that preserves an $U(1)^2$
flavor symmetry and complete non-chirality.
\begin{figure}\begin{center}
\epsfig{file=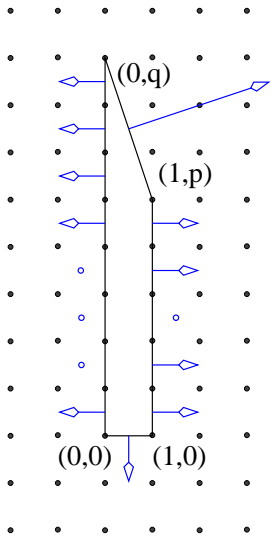, height=7cm}
\caption{Toric diagrams inside the strip.}
\label{strip}\end{center}
\end{figure}
In these cases, instead of having six types of
fields as for a general $\Lpqr$, there are only five types, with multiplicities $p$, $q$,
$q-p$. The charges are given in the table \ref{chargespqp}. As it can
be seen from the toric diagrams, there are parallel external
$(p,q)$-legs. This implies (see \cite{Franco:2005fd,
  Berenstein:2005xa, franco2005} for recent discussions) that there are
\emph{toric} complex structure deformations of these models, as is the
case of the conifold \cite{KlebanovProgram}, which is also
$L^{1,1|1}$. This corresponds to the possibility of pulling out one or
more branes from the $(p,q)$-web.

\begin{table}[!h]\begin{center}
$$\begin{array}{|c|c|c|c|c|} \hline
\hs\mathrm{Field}\hs&\hs\mathrm{number}\hs&\hs Q_R &
\hs Q_H \hs & \hs Q_V \hs\\ \hline\hline
L & p & \hs \frac{2q-p-\sqrt{p^2-pq+q^2}}{3(q-p)} \hs &- 1 & 0\\ \hline
R & q & \frac{q-2p+\sqrt{p^2-pq+q^2}}{3(q-p)} &+ 1 & 0\\\hline
U & p & \frac{2q-p-\sqrt{p^2-pq+q^2}}{3(q-p)} & 0 & +1\\ \hline
D & q & \frac{q-2p+\sqrt{p^2-pq+q^2}}{3(q-p)} & 0 & -1 \\ \hline
A &q-p& \frac{4q-2p-2\sqrt{p^2-pq+q^2}}{3(q-p)} &- 1 & +1\\ \hline\hline
\cO_{LD} & 1 & \frac{q+p+\sqrt{p^2-pq+q^2}}{3} & - p & - q \\\hline
\cO_{RU} & 1 & \frac{q+p+\sqrt{p^2-pq+q^2}}{3} & + q & + p \\\hline
\cO_{LU} & q &\frac{4q-2p-2\sqrt{p^2-pq+q^2}}{3(q-p)} & - 1 & + 1 \\\hline
\cO_{RD} & p &\frac{2q-4p+2\sqrt{p^2-pq+q^2}}{3(q-p)} & + 1 & - 1 \\\hline

\end{array}$$
\caption{Charges of the fields for the $L^{p,q|q}$ quivers.}\label{chargespqp}
\end{center}\end{table}

In addition to the two toric flavor symmetries, there are $p+q-1$ baryonic symmetries. These symmetries,
together with the combination $J_H + J_V$ of flavor symmetries, are
not to be taken into consideration in performing $a$-maximization. The
reason is that the quivers we are considering are completely non
chiral. This means that the bifundamentals are either adjoint fields
or come in pairs. Every pair contains two bifundamentals with opposite
gauge quantum numbers. Since also the interactions are non chiral,
it is clear that the two bifundamentals of every pair have they same
$r$-charge. We can thus impose this equality before doing the
maximization. In our case we see that $L$-fields are in the same pair
of the $U$-fields. The other pair is $D$- and $R$-fields. One is thus
left to maximize a one parameter family of symmetries, that can be taken
to be $R_0 + \lambda (J_H - J_V)$. The results have thus to be quadratic
irrational R-charges, even if in this case there is no non-Abelian symmetry.

The final results for the R-charges is given in table
\ref{chargespqp}. The central charges
is, from the formulas of \cite{Anselmi:1997am}

\beq
c = a = \frac{9}{32} tr (R^3) = \frac
{2 p^3 - 3 p^2 q - 3 p q^2 + 2 q^3 + 2 \sqrt{( p^2-pq+q^2 )^3}}{16 (q - p)^2}
\eeq

 The AdS/CFT formula \cite{malda,Henningson:1998gx} $a = \frac{\pi^3}{4 V}$ relates this central charge
to the volume of the $L^{p,q|q}$ manifold. The result for the volumes
is given in \cite{Cvetic:2005ft} in terms of the solutions of a quartic
equation. It is easy to see that such equation becomes quadratic in the case
$p=s$, $q=r$ and the positive solution matches precisely the field theory result.

It is important that the results obtained so far in this section can
be applied to different quivers, if $q$ and $p$ are not coprime
integers. This corresponds to the possibility of taking different $Z_k$
orbifolds of the same toric Sasaki-Einstein manifold. In this case the
chiral ring generators can be different from the ones we discuss. We thus restrict
to $q$ and $p$ coprime. In this case there is only one quiver, modulo
Seiberg dualities.
\begin{figure}\begin{center}
\epsfig{file=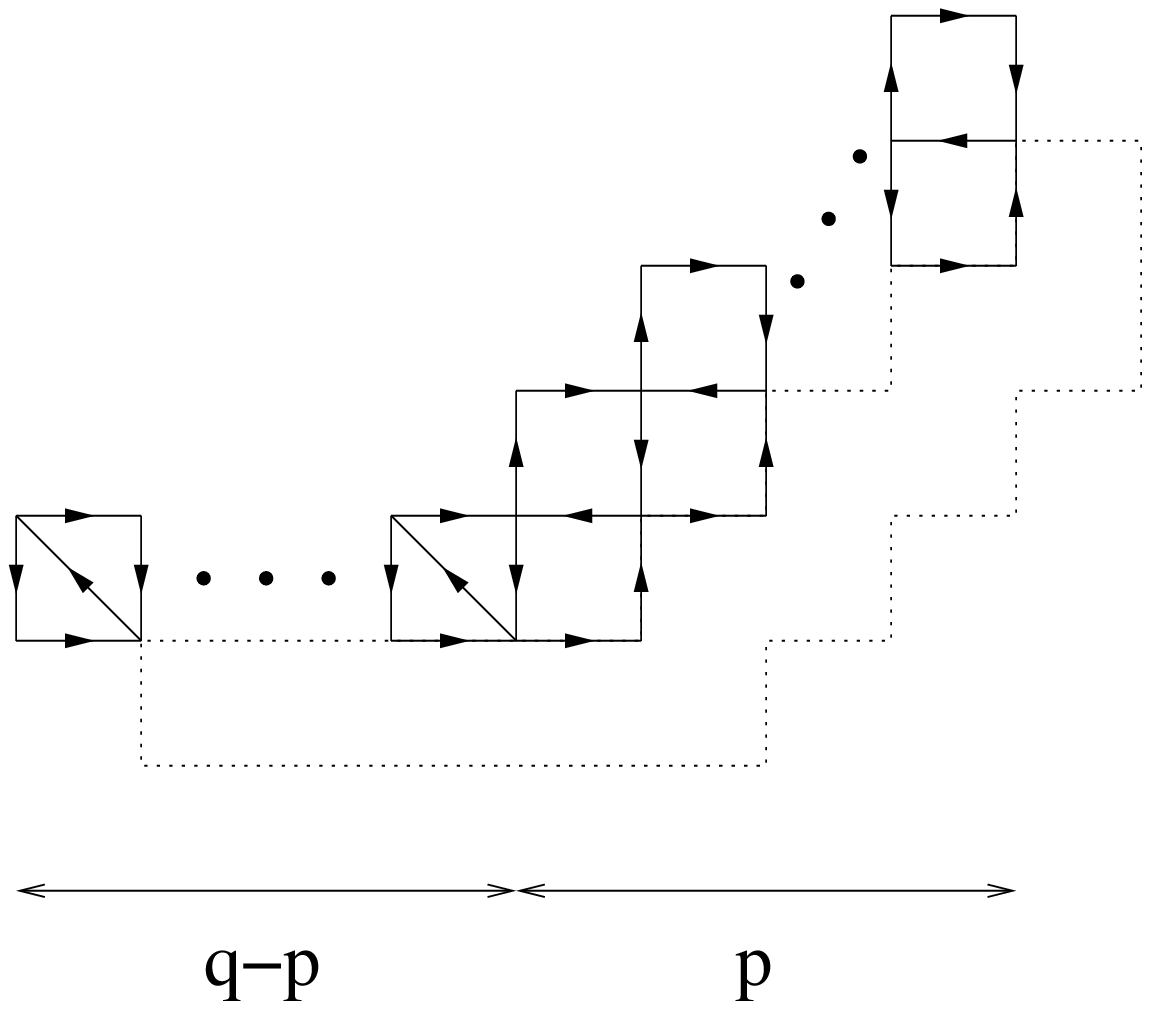, height=7cm}
\caption{A toric phase for the generalized conifold. We show in the
  dotted lines how to glue the fundamental domain.}
\label{pqqquiv}\end{center}
\end{figure}
The chiral ring is generated by four different types of operators. One
subtlety is that some of these operators come with non trivial multiplicities, this
is due to the non smoothness of the background. From the $(p,q)$-web
of $5$ branes (corresponding directly to the toric
diagrams) this fact can be seen as the presence of parallel external
legs. It is the analog of the presence of two $\cL_-$ operators with spin
$0$ in $Y^{p,p}$ \cite{BK}.

There are two long operators, one, $\cO_{LD}$, made of $p$ $L$-fields
and $q$ $D$--fields and one, $\cO_{RU}$, made of $p$ $U$-fields and
$q$ $R$-fields. There are $q-p$ length $1$ operators made with
the $q-p$ different adjoint $A$-fields, that come together with the
$p$ mesons of the form $tr (U L)$. Finally, we find $p$ length $2$ of the form $tr (D R)$.

Now we redefine the two flavor charges $Q_H$ and $Q_V$ in such a way that for each
meson one of the two new charges, that we  call $Q_{\phi}$ and $Q_{\psi}$,
satisfy $Q_{\phi} = - Q_R$ or $Q_{\psi} = - Q_R$. We display the
charges in table \ref{mesonchargespqp2}.
\begin{table}[!h]
\begin{center}
$$\begin{array}{|c||c|c||c|c|} \hline
\hs\mathrm{Meson}\hs& \hs Q_H \hs &\hs Q_V \hs&\hs Q_{\phi}/Q_R \hs &\hs Q_{\psi}/Q_R \hs\\ \hline\hline
\cO_{LD} & - p & - q & - 1 & \frac{p+2q-2\sqrt{p^2-pq+q^2}}{p}\\\hline
\cO_{RU} & + q & + p& \frac{2p+q-2\sqrt{p^2-pq+q^2}}{q} & - 1 \\\hline
\cO_{LU} & - 1 & + 1 & \frac{-p+q+\sqrt{p^2-pq+q^2}}{q}& - 1\\\hline
\cO_{RD} & + 1 & - 1 & - 1 & \frac{p-q+\sqrt{p^2-pq+q^2}}{p}\\\hline
\end{array}$$
\caption{Charge assignments for the four extremal mesonic fields.}
\label{mesonchargespqp2}
\end{center}
\end{table}
This gauge theoretical table is meant to be compared with the geometrical table \ref{mesoncharges} of
section \ref{strings}. By construction the $(-1)$s in the last two
columns match. We now compute the other four values from the geometry.

\subsection{Comparison with the geometry}
In the case of $r=q$ the geometrical formulas of the Appendix simplify.
Eqs. (\ref{chi1}) and (\ref{chi2}) become
\beqa
0 &=& p - 4 q \chi_1 - 4 ( p - q ) \chi_1^2 \\
& & \chi_1 + \chi_2 = \half \label{xi1+xi2=1/2}
\eeqa
which give
\beqa
\chi_1 &=& \frac{-p+\sqrt{p^2-pq+q^2}}{2(q-p)} \\
\chi_2 &=& \frac{q-\sqrt{p^2-pq+q^2}}{2(q-p)}
\eeqa
Due to eq. (\ref{xi1+xi2=1/2}), the square root in eq. (\ref{x1}) simplifies to
\beq
\sqrt{1 - 2 (\chi_1 + \chi_2 ) + ( \chi_2 - \chi_1 )^2} = \chi_2 - \chi_1 \geq 0
\eeq
We thus find
\beqa
\frac{2}{3} - \frac{x_1}{\a} &=& \frac{ p + 2 q - 2\sqrt{p^2-pq+q^2}}{3 p} \\
\frac{2}{3} - \frac{x_2}{\a} &=& \frac{ p - q + \sqrt{p^2-pq+q^2}}{3 p} \\
\frac{2}{3} - \frac{x_1}{\b} &=& \frac{ q - p + \sqrt{p^2-pq+q^2}}{3 q} \\
\frac{2}{3} - \frac{x_2}{\b} &=& \frac{ q + 2 p - 2 \sqrt{p^2-pq+q^2}}{3 q} 
\eeqa

These are precisely the values found on the gauge side, reported in
  Table \ref{mesonchargespqp2}.

%\subsection{Non conformal toric deformations}

%%%%%%%%%%%%%%%%%%%%%%%%%%%%%%%%%%%%%%%%%%%%%%%%%%%%%%%%%%%%%%%%%%%%%%%%%%%%%
\section{General case: $\Lpqr$} \label{generic}

\subsection{R-charges}\label{rcharg}
In this section we perform $a$-maximization \cite{intriligator03} to obtain the R-charges and
compare to the previous results. As a check we also compare the central charge we obtain to
the volume of the Sasaki-Einstein manifold.

Given the charge assignments of table \ref{charges1} and eq.(\ref{Rcharges}), the $a$-function that we should maximize can be written as
\beqa
\tr((R-1)^3) =  \frac{32}{9}a &=& p + q + q x^3 + p (z-x)^3 + s (y-z-1)^3  \nonumber \\
&& + r (-y-1)^3 + (q-s) (y-x)^3 + (q-r) (z-x-y)^3
\eeqa
where we included the contribution $p+q$ from the gauginos.
Now we have to find the point at which $\partial_x a = \partial_y a =
\partial_z a =0$. It is useful to introduce two new variables $\xi_1=x/z$ and $\xi_2 = y/z$.
The first equations to solve implies
\beq
\frac{\partial a}{\partial x} = 0 \ \ \ \ \Longleftrightarrow \ \ \ \ \
\xi_1= \half \frac{s+2(r-q)\xi_2 + (q-p)\xi_2^2}{s+(r-s)\xi_2}
\label{eqxi1}
\eeq
Then we get
\beqa
&& s\frac{\partial a}{\partial y} + (s-r) \frac{\partial a}{\partial z} = 0 \ \ \ \ \Longleftrightarrow \ \ \ \ \ \\
&& \ \ \ z = \frac{2rs(1-2\xi_2)}{p(s-r)\xi_2^2 + 2\xi_1\xi_2(q(r+s)-s^2-r^2)+2pr\xi_2 + 2 s (p-r)\xi_1-sp}
\label{eqz}
\eeqa
 Finally, replacing the expressions for $z$ and $\xi_1$ in the equation $\frac{\partial a}{\partial z} = 0$ we
get a quartic equation for $\xi_2$:
\beq
P_{[4]}(\xi_2)= 0
\label{eqxi2}
\eeq
where $P_{[4]}$ is a polynomial of order four given by
\beqa \label{P4def}
P_{[4]}(\xi_2) &=& 4 (-4r_-^2 p_+^2 + p_+^2p_-^2 + 3 r_-^4)\ \xi_2^4 \\
&& + (32 r_-^2 p_+^2+4r_- p_+ p_-^2+12 r_-^3 p_+-24r_-^4-8p_+^2p_-^2-16 r_- p_+^3)\ \xi_2^3 \nonumber \\
&& +(r_-^2 p_-^2+19 r_-^4-21r_-^2p_+^2-6r_-p_+ p_-^2-18 r_-^3 p_+-4 p_+^4+5 p_+^2 p_-^2+24 r_- p_+^3)\ \xi_2^2 \nonumber\\
&& +(10 r_-^3 p_+-r_-^2 p_-^2-p_+^2 p_-^2-7 r_-^4-12 r_- p_+^3+5 r_-^2 p_+^2+4 p_+^4+2 r_- p_+ p_-^2)\ \xi_2 \nonumber\\
&& +r_-^4-p_+^4+2 r_- p_+^3-2 r_-^3 p_+ \nonumber
\eeqa
where, for brevity we defined $p_{\pm} = p\pm q$ and $r_- = r-s$. When $r=s$ or $p=r$ corresponding to $r_-=0$ or $r_-=p_-$,
the equation factorizes. In the $r_-=0$ case, which corresponds to $Y^{\half p_+,-\half p_-}$, a solution is $\xi_2=\half$
which can be seen to agree with the known result.

\subsection{The volume of the manifold}
 A way to check the $a$-maximization we performed is to compute $a$ and compare with the volume of the $\Lpqr$
manifold. The value of $a$ at the local maximum can be seen to be\footnote{This result is obtained by computing
$\tilde{a}= a - \frac{1}{3}(x\partial_x a+y\partial_y a+z\partial_z a)$ which at the extremes agrees with $a$.}
\beq
 \frac{32}{9} \ \bar{a} = r + s - r ( 1 + \bar{y} )^2 - s ( 1 -
 \bar{y} + \bar{z} )^2
\eeq
where the bars indicate quantities evaluated at the local maximum. We can now obtain an expression in terms
of $\xi_2$:
\beqa
\bar{a} &=& - 18 \,p\,q\,rs\,
  \frac{\xi_2(\xi_2-1)(2\xi_2-1)[p_++r_-(2\xi_2-1)]^2[r_-+p_+(2\xi_2-1)]}{ P(\xi_2)^{2}} \\ 
P(\xi_2) &=& 4p_+(r_-^2-p_-^2)\,\xi_2^3+2\left[2r_-^3-r_-(p_+^2+p_-^2)-3p_+(r_-^2-p_-^2)\right]\,\xi_2^2 \\
           && +2(p_+-r_-)(2r_-^2-p_-^2-p_+^2)\,\xi_2+(p_++r_-)(p_+-r_-)^2
\label{axi2}
\eeqa
 We can rewrite $\bar{a}$ in terms of a variable $W$ as
\beq
\bar{a} = \frac{1}{4} \frac{8pqrs}{(p+q)^3} \frac{1}{W}
\eeq
Since $\xi_2$ obeys the quartic equation (\ref{eqxi2}) that implies that $W$ also satisfies a similar equation.
Using a computer algebra program (\eg\ Maple or Mathematica), it is easy to check that the equation $W$ satisfies
is\footnote{To check this, one replaces (\ref{axi2}) in this equation. After
taking common denominator the numerator can be seen to factorize into $P_{[4]}(\xi_2)$ and a polynomial of order twenty.
Since, by (\ref{eqxi2}), $P_{[4]}(\xi_2)$=0, the equation is satisfied.}:
\beqa
0 &=& (1-f^2)(1-g^2) h_-^4 + 2h_-^2\left[2\left(2-h_+\right)^2 - 3h_-^2\right] W \\
 && \left[8h_+\left(2-h_+\right)^2 - h_-^2(30+9h_+)\right] W^2 \\
 && + 6(2-9h_+)W^3 - 27 W^4
\eeqa
where $f=-p_-/p_+$, $g=r_-/p_+$ and $h_\pm = f^2\pm g^2$. This equation is precisely the same equation
that appeared in \cite{Cvetic:2005ft} and determines the volume $V=\pi^3(p+q)^3W/(8pqrs)$ of the $\Lpqr$
manifold. This implies that the AdS/CFT relation \cite{malda,Henningson:1998gx}
\beq
 a = \frac{\pi^3}{4\mbox{V}}
\eeq
between $a$ and the volume $V$ of the manifold is exactly satisfied. It is perhaps interesting that even if
only one solution of the quartic equation for the volume is physical all solutions actually match. This suggest
that it might be possible to do a more direct derivation of the equivalence between the supergravity
and field theory computations.

\subsection{BPS operators and massless geodesics}
 Now we would like to compare the flavor and R-charges of the operators that we associate with the
geodesics at the ``corners'' of the geometry and that we summarized in table \ref{mesoncharges}.
 The charges of the corresponding operators are summarized in table \ref{mcharges1}. One thing to note is that
since $R_\dla=R_\da+R_\la$ and $R_\ula=R_\ua+R\la$ then the $R$-charges of these particular operators depend
only on their total $Q_H$ and $Q_V$ charges. For other operators that is not the case, for example there are
operators with large R-charge and $Q_H=Q_V=0$, that arise taking
  powers of one basic operator $\mathcal{O}_\b$ with $Q_R=2$ and
$Q_H=Q_V=0$. This short BPS meson $\mathcal{O}_\b$ generates the $\beta$-deformation \cite{Benvenuti:2005wi,
  Lunin:2005jy} and exists for any toric superconformal quiver \cite{Benvenuti:2005wi}. 

 Our first task is to relate the $U(1)$ flavor charges $Q_V$ and $Q_H$ with the isometries $Q_\phi$ and $Q_\psi$
of the background. We found that we obtain a correct matching if we define
\beqa
 Q_V &=& \frac{1}{2} \left( Q_\psi - Q_\phi \right) \\
 Q_H &=& - \frac{\left(A Q_\phi + B Q_\psi \right)}{2 \, p \, q \, (x_2-x_1)} \\
 A &=& ps(\alpha-x_1) + qs(\alpha-x_2) \\
 B &=& pr(\beta-x_1) + qr(\beta-x_2)
\eeqa
The R-charges of the operators can be computed from
eqn. (\ref{Rcharges}) (see table \ref{mcharges1}), with the result
\beqa
R_{LD} &=& -q \yb + s(1-\xb+\zb) \\
R_{RU} &=& r(1+\xb)+p(\yb-\zb) \\
R_{LU} &=& r (1-\xb+\zb) +q (\yb-\zb) \\
R_{RD} &=& s(1+\xb) -p\yb
\eeqa
We remind the reader that $(\xb,\yb,\zb)$ indicates $(x,y,z)$ evaluated at the
local maximum of the central charge $a$.
For the ratios $Q_V/Q_R$ and $Q_H/Q_R$ to match between field theory and supergravity background we need that
\beqa
-\frac{3}{2} \left(1-\frac{x_1}{\alpha}\right) &=& \frac{-q}{-q\yb+s(1-\xb+\zb)} \\
 \frac{3}{2} \left(1-\frac{x_2}{\beta}\right) &=& \frac{p}{r(1+\xb)+p(\yb-\zb)} \\
 \frac{3}{2} \left(1-\frac{x_1}{\beta}\right) &=& \frac{q}{r(1-\xb+\zb)+q(\yb-\zb)} \\
-\frac{3}{2} \left(1-\frac{x_2}{\alpha}\right) &=& \frac{-p}{-p\yb+s(1+\xb)} 
\eeqa
 In the appendix we show that these relations are exactly valid. In fact, together with a further
relation
\beqa
\frac{x_3}{\alpha} &=& -\frac{2}{3\yb} \\
\frac{x_3}{\beta} &=& \frac{2}{3(\yb-\zb)}
\eeqa
can be used to compute all parameters of the geometry in terms of
field theory quantities. We emphasize that these simple relations were
found thanks to the method of comparing massless geodesics with BPS operators.

\section{Conclusions} \label{conclu}
In the present paper we consider massless geodesics moving in the recently found $\Lpqr$
backgrounds. The study of the geodesics give considerable information about the field theory, in particular
they determine a set of four operators which have maximal charges (in modulus) for given $R$-charge. These
are operators which are constituted by elementary fields all with the
same sign of each charge. We find four of them, in correspondence with
the signs of the two flavor charges. On the other hand
an analysis of the toric diagrams of the theories and comparison with the previously known $\cY^{p,q}$ case
suggest a generic construction of the toric representation of the quiver.
 This allows us to conjecture the generic superconformal theories dual to the $\Lpqr$ manifolds. For
those theories we compute the $R$-charges using $a$-maximization and find that the result precisely matches
the computation done in the geometry. In particular a precise mapping is found between the parameters of the
geometry and those that arise in the field theory when performing $a$-maximization. The analysis is
straight-forward albeit cumbersome. For that reason we choose an example of interest, the so called
``generalized conifolds'' which can be identified with $L^{p,q|q}$. In that case we compute explicitly all
the R-charges. In the generic case the results are written in terms of the solutions of a quartic equation on
both sides of the correspondence. The agreement is shown by verifying that the solutions on one side satisfy
the equations on the other side of the correspondence.

In further work, it would be interesting to do a study of extended
semiclassical strings \cite{bmn} in these backgrounds, as was done in
\cite{BK} for the $\cY^{p,q}$ case.

It would also be interesting to see if there is a way of
finding the properties of geodesics starting directly from the toric
diagrams. A similar understanding has been achieved for the volumes in
\cite{Martelli:2005tp}.

%%%%%%%%%%%%%%%%%%%%%%%%%%%%%%%%%%%%
\acknowledgments
We would like to thank Amihay Hanany for many interesting and
enjoyable conversations. We are also grateful to Sebastian Franco,
Pavlos Kazakopoulos, Matt Strassler, David Vegh, Brian Wecht and
Alberto Zaffaroni for discussions.
S.B. has benefited of the warm hospitality of MIT while this work was
being done. M.K. wants to thank the University of Washington for
hospitality while part of this work was being done. The work of M.~K. is supported in part by NSF through
grants PHY-0331516, PHY99-73935 and DOE under grant DE-FG02-92ER40706.

%%%%%%%%%%%%%%%%%%%%%%%%%%%

\section{Useful formulas}

\subsection{The geometry}

 The manifold $\Lpqr$ is defined in terms of two different sets of parameters. One is ($\alpha$, $\beta$) and
the roots $x_1<x_2<x_3$ of the cubic equation $x(\alpha-x)(\beta-x)=\mu$. The other are the integers $p$,$q$,$r$.
 Here we are interested in an explicit relation between the two sets that we derive following \cite{Cvetic:2005ft}.

 The roots $x_{1,2,3}$ satisfy (we assume from now on $\mu=1$)
\beq
x_1 x_2 x_3 = 1, \ \ \ x_1 x_2 +x_1 x_3 + x_2 x_3 = \alpha\beta, \ \ \ x_1+x_2+x_3 = \alpha+\beta
\label{xis}
\eeq
From \cite{Cvetic:2005ft}, the relation to the integers $p$,$q$,$r$ is given through a set of parameters $A_i$, $B_i$ , $C_i$, $i=1,2$ defined
as
\beq
A_i = \frac{\alpha C_i}{x_i-\alpha}, \ \ B_i=\frac{\beta C_i}{x_i-\beta}, \ \ C_i = \frac{(\alpha-x_i)(\beta-x_i)}{2(\alpha+\beta)x_i-\alpha\beta-3x_i^2}
\eeq
and which satisfy
\beq
pC_1 + q C_2=0 , \ \ \ pA_1+qA_2+r=0 , \ \ \ pB_1+qB_2 + s =0
\eeq
Using eq. (\ref{xis}) we can write
\beq
C_1 = -\frac{1}{x_1(x_1-x_2)(x_1-x_3)}, \ \ \ C_2 = -\frac{1}{x_2(x_2-x_1)(x_2-x_3)}
\eeq
We can derive now two equations relating the $x_i$'s to the integers $p$,$q$,$r$:
\beq
 px_2 (x_2-x_3) = q x_1(x_1-x_3), \ \ \ x_1 x_2 +x_1 x_3 +x_2 x_3 = \frac{rs}{pq}(x_1-x_3)(x_2-x_3)
\eeq
which together with $x_1x_2x_3=1$ completely determine $x_i$ in terms of $p$,$q$,$r$. To solve these equations we introduce the
ratios
\beq
\chi_1 = \frac{x_1}{x_3} , \ \ \chi_2= \frac{x_2}{x_3}
\eeq
The equations now reduce to
\beq
p\chi_2(1-\chi_2) = q \chi_1(1-\chi_1), \ \ \ \ \chi_1\chi_2+\chi_1+\chi_2 = \frac{rs}{pq}(1-\chi_1)(1-\chi_2)
\eeq
 The second equation allows to obtain $\chi_2$ as:
\beq\label{chi2}
\chi_2 = \frac{-p\,q\,\chi_1 + rs\,(1-\chi_1)}{p\,q\,(1+\chi_1)\,+\,rs\,(1-\chi_1)}
\eeq
Replacing in the first one, we find a quartic equation for $\chi_1$:
\beqa\label{chi1}
0&=& (p\,q-rs)^2\chi_1^4 + (p\,q-rs)\,(3rs+p\,q)\chi_1^3\\
 & & +(p\,q+rs)\,(3rs-2p^2-p\,q)\chi_1^2 + [p^2(rs-p\,q)-(p\,q+rs)^2]\chi_1 + p^2 rs \nonumber
\eeqa
%\beqa\label{chi1}
%0&=& p^2 r s - ( p^3 q + p^2 q^2 - p^2 r s + 2 p q r s +r^2 s^2 )
%\chi_1 \\
%&& - ( 2 p^3 q + p^2 q^2 + 2 p^2 r s- 2 p q r s - 3 r^2 s^2 ) \chi_1^2 \\
%&& + ( p^2 q^2
% + 2 p q r s - 3 r^2 s^2 ) \chi_1^3 +( p^2 q^2 - 2 p q r s + r^2 s^2 )\chi_1^4
%\eeqaGiven $\chi_1$ we can compute $\chi_2$ as
The other parameters follow trivially as
\beq\label{x1}
x_1 = \left(\frac{\chi_1^2}{\chi_2}\right)^{\frac{1}{3}} , \ \ x_2 = \left(\frac{\chi_2^2}{\chi_1}\right)^{\frac{1}{3}}, \ \ \ x_3 = \frac{1}{(\chi_1\chi_2)^{\frac{1}{3}}}
\eeq
\beqa
\alpha &=& \frac{1+\chi_1+\chi_2+\sqrt{1-2\chi_1+\chi_1^2-2\chi_2+\chi_2^2-2\chi_1\chi_2}}{2(\chi_1\chi_2)^{\frac{1}{3}}} \\
\beta &=& \frac{1+\chi_1+\chi_2-\sqrt{1-2\chi_1+\chi_1^2-2\chi_2+\chi_2^2-2\chi_1\chi_2}}{2(\chi_1\chi_2)^{\frac{1}{3}}}
\eeqa

We finally write
\beqa
\frac{x_i}{\alpha} &=& \frac{2 \chi_i}{1+\chi_1+\chi_2+\sqrt{1-2\chi_1+\chi_1^2-2\chi_2+\chi_2^2-2\chi_1\chi_2}}\\
\frac{x_i}{\beta} &=& \frac{2 \chi_i}{1+\chi_1+\chi_2-\sqrt{1-2\chi_1+\chi_1^2-2\chi_2+\chi_2^2-2\chi_1\chi_2}}
\eeqa
from which it possible to find the value of the $U(1)$-fibration
functions at the four vertices of the coordinate rectangle. Note that,
although we set $\mu = 1$, the results for any ratio of two of the
quantities $x_i, \a, \b$ is independent of $\mu$.

\subsection{Map to the field theory}
 Now we want to find the relation between the parameters $x_{i=1\ldots 3}$, $\alpha$, $\beta$ in the geometry
and those in the field theory. The parameters we consider in the field theory are $x$, $y$, $z$ used in section
\ref{generic}, when performing $a$-maximization. Here we consider them always evaluated at the local maximum in which
case they are functions of $p$, $q$ and $r$ as determined in that section. To emphasize that they are evaluated
at the local maximum we denote them as $\xb$, $\yb$ and $\zb$.

 Analyzing the matching to massless geodesics we were led to certain relations that can be summarized as
follows:
\beqa
\zeta_1 &=& \frac{x_1}{\alpha} =\ 1\ +\ \frac{2}{3}\ \frac{q}{q\yb-s(1-\xb+\zb)} \\
\zeta_2 &=& \frac{x_2}{\alpha} =\ 1\ -\ \frac{2}{3}\ \frac{p}{-p\yb+s(1+\xb)} \\
\zeta_3 &=& \frac{x_3}{\alpha} =\ -\ \frac{2}{3}\ \frac{1}{\yb} \\&& \nonumber\\
\tilde{\zeta}_1 &=& \frac{x_1}{\beta} =\ 1\ -\ \frac{2}{3}\ \frac{q}{q(\yb-\zb)+r(1-\xb+\zb)} \\
\tilde{\zeta}_2 &=& \frac{x_2}{\beta} =\ 1\ -\ \frac{2}{3}\ \frac{p}{p(\yb-\zb)+r(1+\xb)} \\
\tilde{\zeta}_3 &=& \frac{x_3}{\beta} =\ \frac{2}{3}\ \frac{1}{\yb-\zb} 
\eeqa
It is easier, as we did, to write these relations in terms of ratios. In the geometry this amounts to eliminating
the parameter $\mu$. To prove these relations, we proceed to show that they satisfy the same equations
that we derived in the previous subsection (after appropriately dividing by $\alpha$ and eliminating $\beta/\alpha$:
\beqa
&& p\, \zeta_2\, (\zeta_2-\zeta_3) = q\, \zeta_1\, (\zeta_1-\zeta_3) \\
&& \zeta_1\zeta_2 +\zeta_1\zeta_3 +\zeta_2\zeta_3 = \frac{rs}{p\,q} (\zeta_1-\zeta_3)(\zeta_2-\zeta_3) \\
&& \zeta_1+\zeta_2+\zeta_3 = 1 + \zeta_1\zeta_2 +\zeta_1\zeta_3 +\zeta_2\zeta_3
\eeqa
 To check these equations first one replaces $x$, $y$ and $z$ by their expressions in terms of $\xi_2$, namely
following eqs. (\ref{eqxi1}), (\ref{eqz}). After that, the equation becomes a rational function whose numerator is
a polynomial multiple of $P_{[4]}(\xi_2)$ as defined in (\ref{P4def}). Since, at the local extrema, $\xi_2$ is a
root of $P_{[4]}(\xi_2)$, the equations are satisfied.

 These equation completely determine $\zeta_{i=1\ldots 3}$ up to a discrete set of permutations. We checked
using particular examples that the assignments are as we discussed in the case $r>s$ that we are considering.

 The same applies to the $\tilde{\zeta}_{i=1\ldots 3}$. Moreover, as an exercise, one can check that other relations
such as $\zeta_1\tilde{\zeta}_2 - \tilde{\zeta}_1 \zeta_2 =0$ also reduce to zero after using that
$P_{[4]}(\xi_2)=0$.

%%%%%%%%%%%%%%%%%%%%%%%%%%%%%%%%%%%%%%%%%%%%%%%%%%


\begin{thebibliography}{99}

%-----------------------------------------------------------------------------------

\bibitem{malda}
J.~Maldacena,
``The large $N$ limit of superconformal field theories and supergravity,''
Adv.\ Theor.\ Math.\ Phys.\ {\bf 2}, 231 (1998)
[Int.\ J.\ Theor.\ Phys.\ {\bf 38}, 1113 (1998)],
{\tt hep-th/9711200}, \\
%%CITATION = HEP-TH 9711200;%%
%\cite{Gubser:1998bc}
%\bibitem{Gubser:1998bc}
S.~S.~Gubser, I.~R.~Klebanov and A.~M.~Polyakov,
``Gauge theory correlators from non-critical string theory,''
Phys.\ Lett.\ B {\bf 428}, 105 (1998),
[arXiv:hep-th/9802109], \\
%%CITATION = HEP-TH 9802109;%%
%\cite{Witten:1998qj}
%\bibitem{Witten:1998qj}
E.~Witten,
``Anti-de Sitter space and holography,''
Adv.\ Theor.\ Math.\ Phys.\ {\bf 2}, 253 (1998),
[arXiv:hep-th/9802150], \\
%%CITATION = HEP-TH 9802150;%%
%\bibitem{magoo}
O.~Aharony, S.~S.~Gubser, J.~M.~Maldacena, H.~Ooguri and Y.~Oz,
``Large N field theories, string theory and gravity,''
Phys.\ Rept.\ {\bf 323}, 183 (2000)
%[arXiv:hep-th/9905111].
%%CITATION = HEP-TH 9905111;%%

%\cite{kw}
\bibitem{kw}
I.~R.~Klebanov and E.~Witten, ``Superconformal field theory on
three-branes at a Calabi-Yau singularity," Nucl. Phys. B
536:199-218,1998. arXiv:hep-th/9807080
%%CITATION = HEP-TH 9807080;%%

%\cite{Feng:2000mi}
\bibitem{Feng:2000mi}
B.~Feng, A.~Hanany and Y.~H.~He,
``D-brane gauge theories from toric singularities and toric duality,''
Nucl.\ Phys.\ B {\bf 595}, 165 (2001). arXiv:hep-th/0003085.
%%CITATION = HEP-TH 0003085;%%

%\cite{Morrison:1998cs}
\bibitem{Morrison:1998cs}
  D.~R.~Morrison and M.~R.~Plesser,
  ``Non-spherical horizons. I,''
  Adv.\ Theor.\ Math.\ Phys.\  {\bf 3}, 1 (1999)
  [arXiv:hep-th/9810201].
  %%CITATION = HEP-TH 9810201;%%

%\cite{Beasley:1999uz}
\bibitem{Beasley:1999uz}
  C.~Beasley, B.~R.~Greene, C.~I.~Lazaroiu and M.~R.~Plesser,
  ``D3-branes on partial resolutions of abelian quotient singularities of
  Calabi-Yau threefolds,''
  Nucl.\ Phys.\ B {\bf 566}, 599 (2000)
  [arXiv:hep-th/9907186].
  %%CITATION = HEP-TH 9907186;%%

%\cite{Hanany:2001py}
\bibitem{Hanany:2001py}
  A.~Hanany and A.~Iqbal,
  ``Quiver theories from D6-branes via mirror symmetry,''
  JHEP {\bf 0204}, 009 (2002)
  [arXiv:hep-th/0108137].
  %%CITATION = HEP-TH 0108137;%%


%\cite{Beasley:2001zp}
\bibitem{Beasley:2001zp}
  C.~E.~Beasley and M.~R.~Plesser,
  ``Toric duality is Seiberg duality,''
  JHEP {\bf 0112}, 001 (2001)
  [arXiv:hep-th/0109053].
  %%CITATION = HEP-TH 0109053;%%

%\cite{Feng:2000}
\bibitem{Feng:2000}
 B.~Feng, A.~Hanany, Y.~H.~He and A.~M.~Uranga,
  ``Toric duality as Seiberg duality and brane diamonds,''
  JHEP {\bf 0112}, 035 (2001)
  [arXiv:hep-th/0109063].
  %%CITATION = HEP-TH 0109063;%%

%\cite{Feng:2002zw}
\bibitem{Feng:2002zw}
  B.~Feng, S.~Franco, A.~Hanany and Y.~H.~He,
  ``Symmetries of toric duality,''
  JHEP {\bf 0212}, 076 (2002)
  [arXiv:hep-th/0205144].
  %%CITATION = HEP-TH 0205144;%%


%\cite{Feng:2002kk}
\bibitem{Feng:2002kk}
  B.~Feng, A.~Hanany, Y.~H.~He and A.~Iqbal,
  ``Quiver theories, soliton spectra and Picard-Lefschetz transformations,''
  JHEP {\bf 0302}, 056 (2003)
  [arXiv:hep-th/0206152].
  %%CITATION = HEP-TH 0206152;%%

%\cite{Gauntlett:2004zh}
\bibitem{Gauntlett:2004zh}
J.~P.~Gauntlett, D.~Martelli, J.~Sparks and D.~Waldram,
``Supersymmetric AdS(5) solutions of M-theory,''
  Class.\ Quant.\ Grav.\ {\bf 21}, 4335 (2004)
  [arXiv:hep-th/0402153].
  %%CITATION = HEP-TH 0402153;%%

%\cite{Gauntlett:2004yd}
\bibitem{Gauntlett:2004yd}
J.~P.~Gauntlett, D.~Martelli, J.~Sparks and D.~Waldram,
``Sasaki-Einstein metrics on S(2) x S(3),''
arXiv:hep-th/0403002.
%%CITATION = HEP-TH 0403002;%%

%\cite{Gauntlett:2004hh}
\bibitem{Gauntlett:2004hh}
J.~P.~Gauntlett, D.~Martelli, J.~F.~Sparks and D.~Waldram,
``A new infinite class of Sasaki-Einstein manifolds,''
arXiv:hep-th/0403038.
%%CITATION = HEP-TH 0403038;%%

%\cite{Martelli:2004wu}
\bibitem{Martelli:2004wu}
D.~Martelli and J.~Sparks,
``Toric geometry, Sasaki-Einstein manifolds and a new infinite class of AdS/CFT duals,''
arXiv:hep-th/0411238.
%%CITATION = HEP-TH 0411238;%%

%\cite{Cvetic:2005ft}
\bibitem{Cvetic:2005ft}
  M.~Cvetic, H.~Lu, D.~N.~Page and C.~N.~Pope,
  ``New Einstein-Sasaki spaces in five and higher dimensions,''
  arXiv:hep-th/0504225.
  %%CITATION = HEP-TH 0504225;%%

%\cite{MS:2005}
\bibitem{MS:2005}
 D.~Martelli and J.~Sparks,
  ``Toric Sasaki-Einstein metrics on $S^2 \times S^3$,''
  arXiv:hep-th/0505027.
%%CITATION = HEP-TH 0505027;%%

%\cite{ACG:2001}
\bibitem{ACG:2001}
V.~Apostolov, D.~M.~J.~Calderbank and P.~Gauduchon,
 ``The geometry of weakly selfdual Kahler surfaces,''
Compositio Math. 135 (2003) 279-322. arXiv:math.DG/0104233.
%%CITATION = MATH-DG 0104233;%%

%\cite{Benvenuti:2004dy}
\bibitem{Benvenuti:2004dy}
S.~Benvenuti, S.~Franco, A.~Hanany, D.~Martelli and J.~Sparks,
``An infinite family of superconformal quiver gauge theories with Sasaki-Einstein duals,''
arXiv:hep-th/0411264.
%%CITATION = HEP-TH 0411264;%%

%\cite{Bertolini:2004xf}
\bibitem{Bertolini:2004xf}
  M.~Bertolini, F.~Bigazzi and A.~L.~Cotrone,
``New checks and subtleties for AdS/CFT and a-maximization,''
  JHEP {\bf 0412}, 024 (2004)
  [arXiv:hep-th/0411249].
  %%CITATION = HEP-TH 0411249;%%

%\cite{Herzog:2004tr}
\bibitem{Herzog:2004tr}
  C.~P.~Herzog, Q.~J.~Ejaz and I.~R.~Klebanov,
 ``Cascading RG flows from new Sasaki-Einstein manifolds,''
 JHEP {\bf 0502}, 009 (2005)
 [arXiv:hep-th/0412193].
%%CITATION = HEP-TH 0412193;%%

%\cite{Berenstein:2005xa}
\bibitem{Berenstein:2005xa}
  D.~Berenstein, C.~P.~Herzog, P.~Ouyang and S.~Pinansky,
  ``Supersymmetry Breaking from a Calabi-Yau Singularity,''
  arXiv:hep-th/0505029.
  %%CITATION = HEP-TH 0505029;%%

\bibitem{BK}
  S.~Benvenuti and M.~Kruczenski,
  ``Semiclassical strings in Sasaki-Einstein manifolds and long operators in N
  = 1 gauge theories,''
  arXiv:hep-th/0505046.
  %%CITATION = HEP-TH 0505046;%%

%\cite{Mdef}
\bibitem{Mdef}
  R.~G.~Leigh and M.~J.~Strassler,
  ``Exactly marginal operators and duality in four-dimensional N=1 supersymmetric gauge theory,''
  Nucl.\ Phys.\ B {\bf 447}, 95 (1995)  [arXiv:hep-th/9503121].\\
  %%CITATION = HEP-TH 9503121;%%
  B.~Kol,``On conformal deformations,''
  JHEP {\bf 0209}, 046 (2002)
  [arXiv:hep-th/0205141].
  %%CITATION = HEP-TH 0205141;%%

%\cite{Benvenuti:2005wi}
\bibitem{Benvenuti:2005wi}
  S.~Benvenuti and A.~Hanany,
  ``Conformal manifolds for the conifold and other toric field theories,''
  arXiv:hep-th/0502043.
  %%CITATION = HEP-TH 0502043;%%

%\cite{Lunin:2005jy}
\bibitem{Lunin:2005jy}
  O.~Lunin and J.~Maldacena,
  ``Deforming field theories with U(1) x U(1) global symmetry and their gravity
  duals,''
  arXiv:hep-th/0502086.
  %%CITATION = HEP-TH 0502086;%%

%\cite{intriligator03}
\bibitem{intriligator03}
K.~Intriligator and B.~Wecht
``The exact superconformal R symmetry maximizes A,''
Nucl.\ Phys.\ B {\bf 667}, 183 (2003). arXiv:hep-th/0304128.
%%CITATION = HEP-TH 0304128;%%

%\cite{Anselmi:1997am}
\bibitem{Anselmi:1997am}
  D.~Anselmi, D.~Z.~Freedman, M.~T.~Grisaru and A.~A.~Johansen,
  ``Nonperturbative formulas for central functions of supersymmetric
  gauge theories,''
  Nucl.\ Phys.\ B {\bf 526}, 543 (1998)
  [arXiv:hep-th/9708042].\\
  %%CITATION = HEP-TH 9708042;%%
%\cite{Anselmi:1997ys}\bibitem{Anselmi:1997ys}
  D.~Anselmi, J.~Erlich, D.~Z.~Freedman and A.~A.~Johansen,
``Positivity constraints on anomalies in supersymmetric gauge theories,''
  Phys.\ Rev.\ D {\bf 57}, 7570 (1998)
  [arXiv:hep-th/9711035].
  %%CITATION = HEP-TH 9711035;%%

%\cite{Hanany:2005ve}
\bibitem{Hanany:2005ve}
  A.~Hanany and K.~D.~Kennaway,
  ``Dimer models and toric diagrams,''
  arXiv:hep-th/0503149.
%%CITATION = HEP-TH 0503149;%%

%\cite{Hanany:1996ie}
\bibitem{Hanany:1996ie}
A.~Hanany and E.~Witten,
``Type IIB superstrings, BPS monopoles, and three-dimensional gauge dynamics,''
Nucl.\ Phys.\ B {\bf 492}, 152 (1997)
[arXiv:hep-th/9611230].
%%CITATION = HEP-TH 9611230;%%

%\cite{Franco:2005rj}
\bibitem{Franco:2005rj}
  S.~Franco, A.~Hanany, K.~D.~Kennaway, D.~Vegh and B.~Wecht,
  ``Brane dimers and quiver gauge theories,''
  arXiv:hep-th/0504110.
  %%CITATION = HEP-TH 0504110;%%

%\cite{Okounkov:2003sp}
\bibitem{Okounkov:2003sp}
  A.~Okounkov, N.~Reshetikhin and C.~Vafa,
  ``Quantum Calabi-Yau and classical crystals,''
  arXiv:hep-th/0309208.
  %%CITATION = HEP-TH 0309208;%%

%\cite{Hanany:1997tb}
\bibitem{Hanany:1997tb}
  A.~Hanany and A.~Zaffaroni,
  ``On the realization of chiral four-dimensional gauge theories using branes,''
  JHEP {\bf 9805}, 001 (1998) [arXiv:hep-th/9801134].\\
  %%CITATION = HEP-TH 9801134;%%

%\cite{Hanany:1998}
\bibitem{Hanany:1998}
A.~Hanany and A.~M.~Uranga,
  ``Brane boxes and branes on singularities,''
  JHEP {\bf 9805}, 013 (1998)
  [arXiv:hep-th/9805139].\\
  %%CITATION = HEP-TH 9805139;%%

%\cite{GENCON}
\bibitem{GENCON}
 M.~Aganagic, A.~Karch, D.~Lust and A.~Miemiec,
  ``Mirror symmetries for brane configurations and branes at singularities,''
  Nucl.\ Phys.\ B {\bf 569}, 277 (2000)
  [arXiv:hep-th/9903093].\\
  %%CITATION = HEP-TH 9903093;%%

\bibitem{pqwebs}
O.~Aharony and A.~Hanany,
  ``Branes, superpotentials and superconformal fixed points,''
  Nucl.\ Phys.\ B {\bf 504}, 239 (1997)
  [arXiv:hep-th/9704170].\\
  %%CITATION = HEP-TH 9704170;%%
 O.~Aharony, A.~Hanany and B.~Kol,
  ``Webs of (p,q) 5-branes, five dimensional field theories and grid diagrams,''
  JHEP {\bf 9801}, 002 (1998)
  [arXiv:hep-th/9710116].
  %%CITATION = HEP-TH 9710116;%%

%\cite{Uranga:1998vf}
\bibitem{Uranga:1998vf}
  A.~M.~Uranga,
  ``Brane configurations for branes at conifolds,''
  JHEP {\bf 9901}, 022 (1999)
  [arXiv:hep-th/9811004].
  %%CITATION = HEP-TH 9811004;%%

%\cite{Franco:2005sm}
\bibitem{Franco:2005sm}
  S.~Franco, A.~Hanany, D.~Martelli, J.~Sparks, D.~Vegh and B.~Wecht,
  ``Gauge theories from toric geometry and brane tilings,''
  arXiv:hep-th/0505211.
  %%CITATION = HEP-TH 0505211;%%

%\cite{Butti:2005sw}
\bibitem{Butti:2005sw}
  A.~Butti, D.~Forcella and A.~Zaffaroni,
  ``The dual superconformal theory for L(p,q,r) manifolds,''
  arXiv:hep-th/0505220.
  %%CITATION = HEP-TH 0505220;%%

%\cite{HV:2005}
\bibitem{HV:2005}
  A.~Hanany and D.~Vegh, to appear.

%\cite{Benvenuti:2004wx}
\bibitem{Benvenuti:2004wx}
  S.~Benvenuti, A.~Hanany and P.~Kazakopoulos,
  ``The toric phases of the Y(p,q) quivers,''
  arXiv:hep-th/0412279.
  %%CITATION = HEP-TH 0412279;%%

%\cite{Henningson:1998gx}
\bibitem{Henningson:1998gx}
  M.~Henningson and K.~Skenderis,
 ``The holographic Weyl anomaly,''
  JHEP {\bf 9807}, 023 (1998)
  [arXiv:hep-th/9806087].
  %%CITATION = HEP-TH 9806087;%%

\bibitem{KlebanovProgram}
%\cite{Klebanov:1999rd}
%\bibitem{Klebanov:1999rd}
I.~R.~Klebanov and N.~A.~Nekrasov,
``Gravity duals of fractional branes and logarithmic RG flow,''
Nucl.\ Phys.\ B {\bf 574} (2000) 263 [hep-th/9911096];\\
%\cite{Klebanov:2000nc}
%\bibitem{Klebanov:2000nc}
I.~R.~Klebanov and A.~A.~Tseytlin,
``Gravity duals of supersymmetric $SU(N)\times SU(N+M)$ gauge theories,''
Nucl.\ Phys.\ B {\bf 578} (2000) 123 [hep-th/0002159].\\
%\bibitem{KS}
I. R. Klebanov and M. J. Strassler, ``Supergravity and a confining gauge
theory: duality cascades and ${\chi}$SB-resolution of naked singularities,"
JHEP {\bf 0008} (2000) 052.


%\cite{Franco:2005fd}
\bibitem{Franco:2005fd}
  S.~Franco, A.~Hanany and A.~M.~Uranga,
  ``Multi-flux warped throats and cascading gauge theories,''
  arXiv:hep-th/0502113.
  %%CITATION = HEP-TH 0502113;%%

\bibitem{franco2005}
S.~Franco, A.~Hanany, F.~Saad and A.~M.~Uranga,
``Fractional Branes and Dynamical Supersymmetry Breaking,''
arXiv:hep-th/0505040.
  %%CITATION = HEP-TH 0505040;%%

%\cite{bmn}
\bibitem{bmn}
D.~Berenstein, J.~M.~Maldacena and H.~Nastase,
``Strings in flat space and pp waves from N=4 super Yang Mills,''
JHEP {\bf 0204}, 013 (2002) [hep-th/0202021], \\
%%CITATION = HEP-TH 0202021;%%
S.~S.~Gubser, I.~R.~Klebanov and A.~M.~Polyakov,
``A semi-classical limit of the gauge/string correspondence,''
Nucl.\ Phys.\ B {\bf 636}, 99 (2002) [hep-th/0204051], \\
%%CITATION = HEP-TH 0204051;%%
J.~A.~Minahan and K.~Zarembo,
``The Bethe-ansatz for N = 4 super Yang-Mills,''
JHEP {\bf 0303}, 013 (2003) [hep-th/0212208], \\
%%CITATION = HEP-TH 0212208;%%
A.~A.~Tseytlin,
``Spinning strings and AdS/CFT duality,'' hep-th/0311139, \\
%%CITATION = HEP-TH 0311139;%%
M.~Kruczenski,
``Spin chains and string theory,'' hep-th/0311203.
%%CITATION = HEP-TH 0311203;%%

%\cite{Martelli:2005tp}
\bibitem{Martelli:2005tp}
  D.~Martelli, J.~Sparks and S.~T.~Yau,
  ``The geometric dual of a-maximisation for toric Sasaki-Einstein manifolds,''
  arXiv:hep-th/0503183.
%%CITATION = HEP-TH 0503183;%%

\end{thebibliography}
\end{document}